\documentclass[12pt]{article}

\usepackage[dvips]{graphicx}
\usepackage{apalike}

\newcommand{\eeqq}{\end{eqnarray*}}

\begin{document}
\bibliographystyle{apalike}

\title{SAMPLING PROPERTIES OF THE SPECTRUM AND COHERENCY OF SEQUENCES OF ACTION POTENTIALS}
\author{M. R. Jarvis$^{(1)}$ and P. P. Mitra$^{(2)}$}

\date{\today}
\maketitle
\begin{center} 
\it{ $^{(1)}$ Division of Biology, California Institute of Technology, Pasadena, 91125 USA}\\
\it{ $^{(2)}$ Bell Laboratories, Lucent Technologies, Murray Hill, New Jersey, 07974 USA}\\
\end{center}

\begin{abstract}
\noindent
The spectrum and coherency are useful quantities for characterizing the temporal correlations and functional relations within and between point processes.  This paper begins with a review of these quantities, their interpretation and how they may be estimated.  A discussion of how to assess the statistical significance of features in these measures is included.  In addition, new work is presented which builds on the framework established in the review section.  This work investigates how the estimates and their error bars are modified by finite sample sizes.  Finite sample corrections are derived based on a doubly stochastic inhomogeneous Poisson process model in which the rate functions are drawn from a low variance Gaussian process.  It is found that, in contrast to continuous processes, the variance of the estimators cannot be reduced by smoothing beyond a scale which is set by the number of point events in the interval.  Alternatively, the degrees of freedom of the estimators can be thought of as bounded from above by the expected number of point events in the interval.  Further new work describing and illustrating a method for detecting the presence of a line in a point process spectrum is also presented, corresponding to the detection of a periodic modulation of the underlying rate.  This work demonstrates that a known statistical test, applicable to continuous processes,  applies, with little modification, to point process spectra, and is of utility in studying a point process driven by a continuous stimulus.  While the material discussed is of general applicability to point processes attention will be confined to sequences of neuronal action potentials (spike trains) which were the motivation for this work.\\

\end{abstract}

\section{Introduction}
\label{Intro}

The study of spike trains is of central importance to electrophysiology.  Often changes in the mean firing rate are studied but there is increasing interest in characterising the temporal structure of spike trains, and the relationships between spike trains, more completely \cite{s1,gp1,ab1}.  A natural extension to estimating the rate of neuronal firing is to estimate the autocorrelation and the cross-correlation functions\footnote{Definitions of these quantities will be given in section \ref{Defin}.}.  This paper will discuss the frequency domain counterparts of these quantities.  Auto- and cross-correlations correspond to spectra and cross spectra respectively.  The coherency, which is the normalised cross spectrum, does not in general have a simple time domain counterpart.

The frequency domain has several advantages over the time domain.  Firstly often subtle structure can be detected with the frequency domain estimators which is difficult to observe with the time domain estimators.  Secondly, the time domain quantities have problems which are associated with sensitivity of the estimators to weak non-stationarity and the non-local nature of the error bars \cite{cb1}.  These problems are greatly reduced in the frequency domain.  Thirdly, reasonably accurate confidence intervals may be placed on estimates of the second order properties in the frequency domain which permits the statistical significance of features to be assessed.  Fourthly, the coherency provides a normalised measure of correlations between time series, in contrast with time domain cross-correlations which are not normalisable by any simple means.

This paper begins by reviewing the population spectrum and coherency for point processes and motivating their use by describing some example applications.  Next direct, lag window and multitaper estimators of the spectrum and coherency are presented.  The concept of degrees of freedom is introduced and used to obtain large sample error bars for the estimators.   Many elements of the work discussed in the review section of this paper can be found in the references \cite{pw1,cl1,br1,bt1}.  Most of the material in these references is targeted at either spectral analysis of continuous processes or at the analysis of point processes but with less emphasis on spectral analysis.  Building on this framework corrections, based on a specific model, will be given for finite sample sizes.  These corrections are cast in terms of a reduction in the degrees of freedom of the estimators.  For a homogeneous Poisson process the modified degrees of freedom is the harmonic sum of the the asymptotic degrees of freedom and twice the number of spikes used to construct the estimate.  Modifications to this basic result are given for structured spectra and tapered data.   A section is included on the treatment of point process spectra which contain lines.  A statistical test for the presence of a line in a background of coloured noise is given, and the method for removal of such a line described.  An example application to periodic stimulation is given.

\section{Population measures and their interpretation}
\label{Theor}

\subsection{Counting representation of a spike train}
\label{Count}

\begin{figure}[hbtp] 
\begin{center}
\leavevmode
\hbox{%
\unitlength=1mm
\begin{picture}(0,32)
\put(-70,0){\includegraphics[height=2cm]{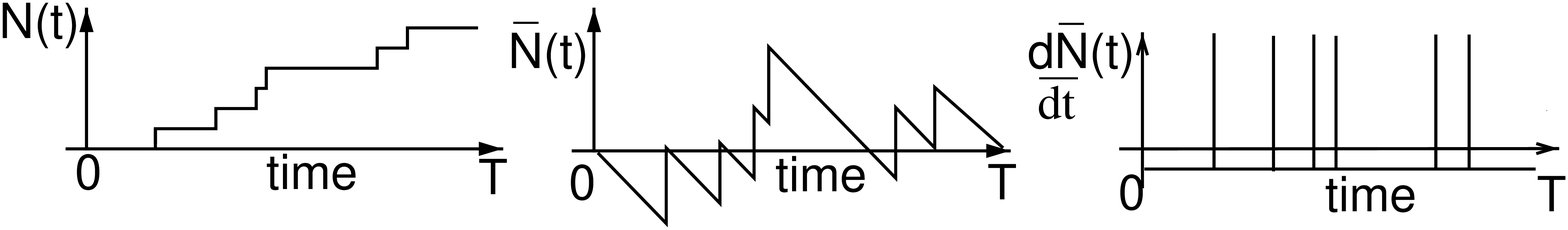}}
\end{picture}}
\end{center}
\caption{Example illustrating how the processes $N$, $\overline{N}$ and $d\overline{N}/dt$ relate to each other.  The vertical lines in the process $d\overline{N}/dt$ depict delta functions.}
\label{N_Nbar_dNbar}
\end{figure}

A spike train may be regarded as a point process.  If the spike shapes are neglected, it is completely specified by a series of spike times $\{t_i\}$ and the start and end points of the recording interval $[0,T]$.  It is convenient to introduce some notation which enables the subsequent formulae to be written in a compact form \cite{br1}.  The counting process $N(t)$ is defined as the number of spikes which occur between the start of the interval $(t=0)$ and time $t$.  The counting process has the property that the area beneath it grows as $t$ becomes larger.  This is undesirable because it leads to an interval dependent peak at low frequencies in the spectrum.  To avoid this a process $\overline{N}(t) = N(t) - \lambda t$, where $\lambda$ is the mean rate, which has zero mean may be constructed.  Note that $d\overline{N}(t) = \overline{N}(t+dt) - \overline{N}(t)$ which is either $1 - \lambda dt$ \hspace{1mm} or $-\lambda dt$ depending on whether or not there is a spike in the interval $dt$.  Thus $d\overline{N}(t)/dt$ is a series of delta functions\footnote{A delta function is a generalized function.  It has an area of one beneath it but has zero width and therefore infinite height.} with the mean rate subtracted.  Figure \ref{N_Nbar_dNbar} illustrates the relationship between $N(t)$, $\overline{N}(t)$, and $d\overline{N}(t)/dt$. 

\subsection{Stationarity}
\label{Stati}

It will be assumed in what follows that the spike trains are second order stationary.  This means that their first and second moments do not depend on the absolute time.  In many electrophysiology experiments this is not the case.  In awake behaving studies the animal is often trained to perform a highly structured task.  Nevertheless it may still be the case that over an appropriately chosen short time window, the statistical properties are changing slowly enough for reasonable estimates of the spectrum and coherency to be obtained.  As an example, neurons in primate parietal area PRR exhibit what is known as memory activity during a delayed reach task \cite{Reach1}.  The mean firing rate of these neurons varies considerably during the task but during the memory period is roughly constant.  The assumption of stationarity during the memory period is equivalent to the intuitive notion that there is nothing special about 0.75s into the memory period as compared to say 0.5s.  Second order stationarity implies that the mean firing rate ($\lambda$) is independent of time and additionally that the autocovariance depends only on the lag ($\tau$) and not on the absolute time.

\subsection{Definitions}
\label{Defin}

Equations \ref{E1} - \ref{E2} give the first and second order moments of a single spike train for a stationary process.  The spectrum $S(f)$ is the Fourier transform of the autocovariance function $(\mu(\tau) + \lambda \delta(\tau))$.  

\begin{eqnarray}
& &\frac{E\{dN(t)\} }{dt} =  \lambda \label{E1} \\
& &\frac{E\{d\overline{N}(t)\}}{dt} =  0 \\
& &\mu(\tau) + \lambda \delta(\tau) = \frac{E[d\overline{N}(t)d\overline{N}(t+\tau)]}{dtd\tau} \label{E1.5} \\
& &S(f) = \lambda + \int_{-\infty}^{\infty} \mu(\tau) \exp(-2\pi i f \tau) d\tau \label{E2}
\end{eqnarray}
Where $E$ denotes the expectation operator.\\

The autocovariance measures how likely it is that a spike will occur at time $t+\tau$ given that one has occurred at time $t$.  Usually $\mu(\tau)$ is estimated rather than the full autocovariance which includes a delta function at zero lag\footnote{When estimating the autocovariance using a histogram method one generally omits the spike at the start of the interval which would always fall in the bin nearest zero.}.  However, in order to take the Fourier transform the full autocovariance is required.  The inclusion of this delta function leads to a constant offset of the spectrum.  This offset is an important difference between continuous time processes and point processes.  The population coherency $\gamma(f)$ is defined in equations \ref{C1} - \ref{C2}.

\begin{eqnarray}
& &\mu_{ab}(\tau) = \frac{E[d\overline{N}_a(t)d\overline{N}_b(t+\tau)]}{dtd\tau} \label{C1} \\
& & S_{ab}(f) = \int_{-\infty}^{\infty} \mu_{ab}(\tau) \exp(-2\pi i f \tau) d\tau + \lambda_a \delta_{ab}\\
& &\gamma(f) = \frac{S_{12}(f)}{\sqrt{S_{11}(f)S_{22}(f)}} \label{C2}
\end{eqnarray}
Where indices $1$ and $2$ denote simultaneously recorded spike trains from different cells.\\

Unlike the spectrum, which is strictly real and positive, the coherency is a complex quantity.  The modulus of the coherency, which is known as the coherence\footnote{Some authors define coherence as the modulus squared of the coherency.}, can only vary between zero and one.  This makes coherence particularly attractive for detecting relationships between spike trains as it is insensitive to the mean spike rates.

\section{Examples and their interpretation}
\label{Examp}

Before discussing the details regarding how to estimate the spectrum and coherency it will be helpful to motivate them further by considering some simple examples.

\subsection{Example population spectra}
\label{thsp}

\begin{figure}[hbtp] 
\begin{center}
\leavevmode
\hbox{%
\unitlength=1mm
\begin{picture}(0,25)
\put(-55,-5){\includegraphics[height=3cm]{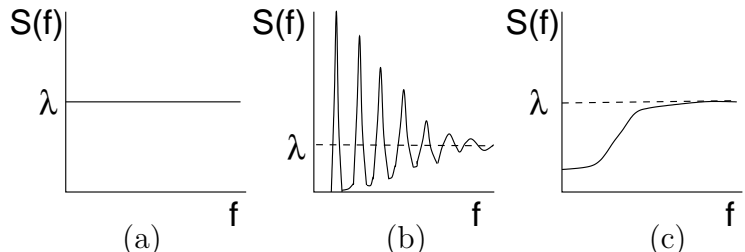}}
\put(-40,-8){(a)} 
\put(-5,-8){(b)} 
\put(30,-8){(c)} 
\end{picture}}
\end{center}
\caption{Example population spectra for different types of underlying process.  (a) Homogeneous Poisson process with rate $\lambda$.  (b) Regularly spaced spikes with jitter.  (c) Spike trains in which the probability of firing is suppressed immediately after firing.} 
\label{specegs}
\end{figure}

For a homogeneous Poisson process of constant rate $\lambda$ the autocovariance is simply $\lambda \delta (\tau)$ and hence the spectrum is a constant equal to the rate $\lambda$.  At the opposite extreme consider the case where the spikes are spaced by intervals $\Delta \tau$.  This is not a stationary process but if a small amount of drift is permitted, so that over an extended period there is nothing special about a given time, it becomes stationary.  The spectrum of this process contains sharp lines at integer multiples of $f = \frac{1}{\Delta \tau}$.  Due to the drift the higher harmonics will become increasingly blurred and in the high frequency limit the spectrum will tend towards a constant value of the mean rate $\lambda$.  As a final example consider the case where $\mu(\tau)$ is a negative Gaussian centered on zero $\tau$.  This form of $\mu(\tau)$ is consistent with the probability of firing being suppressed after firing\footnote{This need not necessarily correspond to the biophysical refractive period but, it could arise, rather from a characteristic integration time.}.  The spectrum of this process will be below $\lambda$ at low frequencies and will go to a constant value $\lambda$ at high frequencies.   Figure \ref{specegs} illustrates these different population spectra.

\subsection{Example population coherency}
\label{thco}

The population coherency of two homogeneous Poisson processes is zero.  In contrast if two spike trains are equal then the coherence is one and the phase of the coherency is zero at all frequencies.  If two spike trains are identical but offset by a lag $\Delta \tau$ then the coherence will again be one but the phase of the coherency will vary linearly with frequency with a slope proportional to $\Delta \tau$ and given by $\phi(f) = 2\pi f \Delta \tau$.

\section{Estimating the spectrum}
\label{estthespec}

Section \ref{Examp} demonstrated that the population spectrum may provide insights into the nature of a spike train.  In this section the question of how to estimate the spectrum from a finite section of data will be introduced.  In what follows the population quantity $\lambda$ in the definition of $\overline{N}(t)$ is replaced by a sample estimate $N(T)/T$.

\subsection{Direct Spectral Estimators}

\subsubsection{Definition}

A popular, though seriously flawed, method for estimating the spectrum is to take the modulus squared of the Fourier transform of the data $d\overline{N}(t)$.  This estimate is known as the Periodogram and is the simplest example of a direct spectral estimator.  More generally, a direct spectral estimator is the modulus squared of the Fourier transform of the data but with the data being multiplied by an envelope function $h(t)$, known as a taper \cite{pw1}.   Equations \ref{D1} - \ref{D3} define the direct estimator.  On substituting $\overline N(t)$ into equation \ref{D2} a form amenable to implementation on a computer is obtained (equation \ref{D5}).  In this form the Fourier transform may be computed rapidly and without the need for the binning of data.   Note that equation \ref{D3} results in $h(t)$ scaling as $1/\sqrt{T}$ as the sample length is altered.  This ensures proper normalization of the Fourier transformation as sample size varies.
\begin{eqnarray}
& & I^{D}(f)  =  | J^{D} (f)|^2 \label{D1} \\
& & J^{D}(f)  =  \int_{0}^{T} h(t) e^{-2 \pi i f t} d\overline{N}(t) \label{D2}
\end{eqnarray}
Where,
\begin{eqnarray}
& & \int_{0}^{T} h(t)^2 dt  =  1 \label{D3} \\
& & J^{D}(f)  =  \sum_{j=1}^{N(T)} h(t_j) e^{-2 \pi i f t_j} - \frac{N(T)H(f)}{T} \label{D5}
\end{eqnarray}
and $H(f)$ is the Fourier transform of the taper.\\

The direct estimator suffers from bias and variance problems, described below, and is of no practical relevance for a single spike train sample. 

\subsubsection{Bias}
\label{bias}

It may not be immediately apparent why the above procedure is an estimate of the spectrum, especially when one is permitted to multiply the data by an arbitrary, albeit normalized, taper.  The relation between $I^D(f)$ and the spectrum may be obtained by taking the expectation of equation \ref{D1}.

\begin{eqnarray}
E\{I^{D}(f)\} = E\{ 
\int_{-\infty}^{\infty} \int_{-\infty}^{\infty} h(t)h(t') e^{-2 \pi i f (t-t')} d\overline{N}(t) d\overline{N}(t') \} \label{F1}
\end{eqnarray}

Assuming that the integration and expectation operations may be interchanged and substituting equation \ref{E1.5} yields \footnote{For the moment, we assume that the population quantity $\lambda$ is known.  This is of course not the case in practice, and one employs the estimate $N(T)/T$ as stated before.  The effect of this extra uncertainty is given in equation \ref{bias2}.},

\begin{equation}
E\{I^{D}(f)\} =  
\int_{-\infty}^{\infty} \int_{-\infty}^{\infty} h(t)h(t') e^{-2 \pi i f (t-t')} \{ \mu(t-t') + \lambda \delta(t-t') \} dt dt'\label{F2}
\end{equation}
Which may be rewritten in the Fourier domain as,
\begin{equation}
E\{I^{D}(f)\} =  \int_{-\infty}^{\infty} S(f')|H(f-f')|^2 df'
\end{equation}

\begin{figure}[hbtp] 
\begin{center}
\leavevmode
\hbox{%
\unitlength=1mm
\begin{picture}(0,25)
\put(-25,-8){\includegraphics[height=3.5cm]{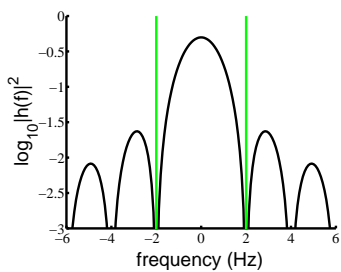}}
\end{picture}}
\end{center}
\caption{The smoothing kernel $|H(f)|^2$.  This is the expected direct estimate of the spectrum in the case of a population spectrum which has a delta function (very sharp feature) at the center frequency.  A rectangular taper of length 0.5s was used.  Solid vertical lines are drawn at $\pm$ the Raleigh frequency.}
\label{biaseg}
\end{figure}

The expected value of the direct estimator is a convolution of the true spectrum and the modulus squared of the Fourier transform of the taper.  The normalization condition on the taper (equation \ref{D3}) is equivalent to the requirement that the kernel of the convolution has unit area underneath it.  Sharp features in the true spectrum will be thus be smeared by an amount which depends on the width of the taper in the frequency domain.  If the taper is well localized in the frequency domain the expected value of the direct estimate is close to the true spectrum but if the taper is poorly localized then the expected value of the direct estimator will be incorrect i.e. \hspace{-1.5mm} the direct spectral estimator is biased.  There is a fundamental level beyond which the bias cannot be reduced, due to the uncertainty relation forbidding simultaneous localization of a function in the time and frequency domains below a given limit.  Since the maximum width of the taper is T the minimum frequency spread is 1/T which is known as the Raleigh frequency.  Figure \ref{biaseg} shows the smoothing kernel for a rectangular taper and a T of 0.5s.  Note that this kernel has large sidelobes which is the primary motivation for using tapering.

In the above argument equation \ref{E1.5} was used in spite of the appearance of the population quantity $\lambda$ rather than the sample estimate $N(T)/T$ for which equation \ref{F1} was defined.  A more careful treatment, which includes this correction, leads to an additional term at finite sample sizes in the expectation of the direct spectral estimator at low frequencies.  The full expression is given below,

\begin{equation}
E\{I^D (f) \} = \int_{-\infty}^{\infty} S(f')|H(f-f')|^2 df' - |H(f)|^2 S(0)/T  \label{bias2}
\end{equation}

In the case of the periodogram, where $h(t) = 1/\sqrt{T}$, the effect is clear since in this case $J^D(0) = 0$ and hence $I^D(0) = 0$ for any set of spike times and any T.  

\subsubsection{Asymptotic variance}

\begin{figure}[hbtp] 
\begin{center}
\leavevmode
\hbox{%
\unitlength=1mm
\begin{picture}(0,25)
\put(-25,-10){\includegraphics[height=3.5cm]{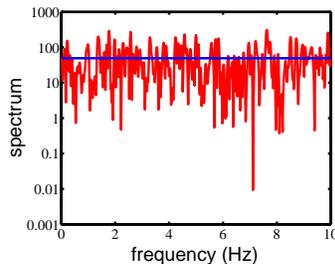}}
\end{picture}}
\end{center}
\caption{An example of a direct spectral estimate.  A 40\% cosine taper was used.  A sample of duration 20s was drawn from a homogeneous Poisson process with a constant rate of 50 Hz.  The population spectrum for this process is flat and is shown by the solid horizontal line.  The direct spectral estimate is clearly noisy although on average the correct spectrum is obtained.}
\label{vareg}
\end{figure}

In the previous section it was shown that provided the taper is sufficiently local in frequency the expected value of the direct spectral estimator will be close to the true spectrum.  However, the fact that the estimate is on average close to the true spectrum belies a serious problem with direct spectral estimators, namely that the estimates have very large fluctuations about this mean.  The underlying source of this problem is that one is attempting to estimate the value of a function at an infinite number of points using a finite sample of data.  The problem manifests itself in the fact that direct spectral estimators are {\it inconsistent} estimators of the spectrum\footnote{Inconsistent estimators have a finite variance even for an infinite length sample.}.  In fact it may be shown that, under fairly general assumptions, the estimates are distributed exponentially (or equivalently as $S(f)\chi_2^2/2$) for asymptotic sample sizes (i.e. \hspace{-1.0mm} $T \rightarrow \infty$) \cite{br72}.  Figure \ref{vareg} illustrates that direct spectral estimators are noisy and untrustworthy, a fact emphasised by the observation that the $\chi_2^2$ distribution has a standard deviation equal to its mean.  In the next three subsections methods for reducing the variance of direct spectral estimators using different forms of averaging will be discussed.  

\subsection{Trial averaging}
\label{trials}

If there are a number of trials ($N_T$) available then the variance of the direct estimator may be reduced by trial averaging. 

\begin{equation}
I^{DT}(f) = \frac{1}{N_T} \sum_{n=1}^{N_T} I_n^D(f) \label{TA1}
\end{equation}
Where $I_n^D(f)$ is the direct spectral estimate based on the $n^{th}$ trial.\\

In the large T limit taking the average entails summing $N_T$ independent samples from a $\chi_2^2$ distribution the result of which is distributed as $\chi_{2N_T}^2$.  The reduction in variance is inversely proportional to the number of trials corresponding to a reduction in standard deviation which is the familiar factor of $1/\sqrt{N_T}$. 

At first sight it appears one would be getting something for nothing by breaking a single section of data into $N_T$ segments and treating them as separate trials.  This is, of course, not the case.  The reason is that if the data is segmented into short length samples, there is loss of frequency resolution proportional to the inverse of segment length.  Lag window and multitaper estimators use the information from these independent estimates without artificially segmenting the data.

\subsection{Lag Window Estimates}

A powerful property of the frequency domain is that, unless two frequencies are very close together, direct estimates of the spectrum of a stationary process at different frequencies are nearly uncorrelated.  This property arises when  the covariance between frequencies falls off rapidly.  If the true spectrum varies slowly over the width of the covariance then the large sample covariance of a direct spectral estimator is given by equation \ref{cov1}.

\begin{equation}
cov\{I^D(f_1),I^D(f_2)\} \simeq E\{I^D( \overline{f})\}^2 \left | \int_{-\infty}^{\infty} h(t)^2 e^{-2 \pi i \Delta f t} dt \right |^2 \label{cov1}
\end{equation}
Where $\overline{f} = (f_1 + f_2) / 2$  and $\Delta f = f_1 - f_2$\\

For $\Delta f = 0$, this expression reduces to the previously mentioned result that the variance of the estimator is equal to the square of the mean.  For $\Delta f >> 1/T$, \linebreak $|\int_{-\infty}^{\infty} h(t)^2 e^{-2\pi i \Delta f t}dt|^2 \rightarrow 0$, since $h(t)^2$ is a smooth function with extent $~T$.  This implies that $cov\{I^D(f_1),I^D(f_2)\} \approx 0$ for $|f_1 - f_2| >> 1/T$.  The approximate independence of nearby points means that, if the true spectrum varies slowly enough, then closely spaced points will provide several independent estimates of the same underlying spectrum.  This is the motivation for the lag window estimator which is simply a smoothed version of the direct spectral estimator \cite{pw1}.  The lag window estimator is defined in equations \ref{LW1} and \ref{LW2}.

\begin{equation}
I^{LW}(f) = \int_{-\infty}^{\infty} K(f-f')I^D(f')df' \label{LW1}
\end{equation}
Where,
\begin{equation}
\int_{-\infty}^{\infty} K(f)df = 1 \label{LW2}
\end{equation}

Averaging over trials may be included by using the trial averaged direct spectral estimate $I^{DT}$ (see equation \ref{TA1}) in place of $I^{D}$ in the above expression.  It is assumed that $K(f)$ is a smoothing kernel with reasonable properties. 

\subsubsection{Bias}

The additional smoothing of the lag window kernel modifies the bias properties of the estimator from those expressed in equation \ref{bias2}.  The expected value of the lag window estimator is given by,

\begin{equation}
E\{ I^{LW}(f) \} = \int_{-\infty}^{\infty} K(f-f')|H(f'-f'')|^2S(f'') df'df'' - \frac{S(0)}{T} \int_{-\infty}^{\infty} K(f-f')|H(f')|^2 df' \label{mtbias}
\end{equation}

\subsubsection{Asymptotic Variance}

The large sample variance of this estimator is readily obtained using equation \ref{cov1}.

\begin{equation}
var\{I^{LW}(f)\}  = \frac{\xi}{N_T} E\{I^{LW}(f)\}^2 \label{trav}
\end{equation}
Where,
\begin{equation}
\xi  = \int_{-\infty}^{\infty} \int_{-\infty}^{\infty} K(f)K(f')|{\cal H}(f-f')|^2 df df' \label{mtvar}
\end{equation}
and,
\begin{equation}
{\cal H}(f)  =  \int_{-\infty}^{\infty} h(t)^2 e^{-2 \pi i f t} dt
\end{equation}

Equation \ref{trav} includes the reduction in variance due to trial averaging.  $1/\xi$ can be interpreted as the effective number of independent estimates beneath the smoothing kernel, as demonstrated by the following qualitative argument.  If $\Delta f$ is the frequency width of the smoothing kernel $K(f)$ and $\delta f$ is the frequency width of the taper ${\cal H}(f)$ then since $K(f) \sim 1/\Delta f$ it follows that $\xi \sim 1/(\Delta f)^2 \int_{\Delta f} \int_{\Delta f} |{\cal H}(f-f')|^2 df df'$ and hence that $\xi \sim  \delta f/\Delta f.$ 

\subsection{Multitaper Estimates}

While the lag window estimator is based on the idea that nearby frequencies provide independent estimates, the estimation is not very systematic, since, one should be able to explicity decorrelate nearby frequencies from the knowledge of the correlations introduced by a finite window size.  This is achieved in multitaper spectral estimation.  The basic idea of multitaper spectral estimation is to average the spectral estimates from several orthogonal tapers.  The orthogonality of the tapers ensures that the estimates are uncorrelated for large samples (consider substituting $h_1(t)h_2(t)$ for $h(t)^2$ in equation \ref{cov1}).  A critical question is the choice of a set of orthogonal tapers.  A natural choice are the discrete prolate spheroidal sequences (dpss) or Slepian sequences, which are defined by the property that they are maximally localised in frequency.  The dpss tapers maximize the spectral concentration defined as;  
  
\begin{equation}
\lambda = \frac{\int_{-W}^{W} |H(f)|^2 df}{\int_{-\infty}^{\infty}  |H(f)|^2 df} \label{conc1}
\end{equation}
Where in the time domain $h(t)$ is strictly confined to the interval [0,T].\\

For given values of W and T there are a finite number of tapers which have concentrations ($\lambda$) close to one, and therefore have well controlled bias.  This number is known as the Shannon number and is $2WT$.  This sets an upper limit on the number of independent estimates that can be obtained for a given amount of spectral smoothing. 

A direct multitaper estimate of the spectrum is defined in equation \ref{MT1}.

\begin{equation}
I^{MT}(f) = \frac{1}{K} \sum_{k=0}^{K-1} I_k^D(f) \label{MT1}
\end{equation}

The eigenspectra $I_k^D$ are direct spectral estimates based on tapering the data with the $k^{th}$ dpss function.  As previously trial averaging can be included by using $I^{DT}$ rather than $I^D$.  More sophisticated estimates involve adaptive (rather than constant) weighting of the data tapers \cite{pw1}.  Multitaper spectral estimation has been recently shown to be useful for analysing neurobiological time series, both continuous processes \cite{bp1} and spike trains \cite{bp2}.

\subsubsection{Bias}

The bias for the multitaper estimate is given by equation \ref{bias2} but with $|H(\cdot)|^2$ replaced by an average over tapers $\frac{1}{K} \sum_{k=0}^{K-1} |H_k(\cdot)|^2$.

\subsubsection{Asymptotic Variance}

The asymptotic variance of the multitaper estimator, including trial averaging, is given by equation \ref{MT2}.

\begin{equation}
var\{I^{MT}(f)\} = \frac{1}{N_TK}E\{I^{MT}(f)\}^2 \label{MT2}
\end{equation}

\subsection{Degrees of freedom}

At this point it is useful to introduce the concept of the degrees of freedom ($\nu_0$) of an estimate.  The degrees of freedom is twice the number of independent estimates of the spectrum.  Degrees of freedom is a useful concept as it permits the expressions for the variance of the different estimators to be written in a common format.

\begin{equation}
var\{I^{X}(f)\} = \frac{2 E\{I^X(f) \}^2}{\nu_0} \label{dof1} 
\end{equation}
Where,
\begin{center}
\begin{tabular}{|p{1.0cm}||p{2.0cm}|p{2.0cm}|p{2.0cm}|p{2.0cm}|} \hline
$X$   & $D$  & $DT$   & $LW$          & $MT$    \\ \hline
$\nu_0$ & $2$  & $2 N_T$ & $2 N_T/\xi$ & $2 N_TK$  \\ \hline
\end{tabular} 
\end{center}

Degrees of freedom is also a useful framework in which to cast both finite size corrections and the confidence limits for the spectra and coherence.  

The variance of estimators of the spectrum can be estimated using internal methods such as the bootstrap or jackknife \cite{BE1},\cite{dt2}.  Jackknife estimates can be constructed over trials or over tapers.  If $\nu_0$ is large ($>20$), then the theoretical and Jackknife variance are in close agreement.  If distributional assumptions can be validly made about the point process, theoretical error bars have an important advantage over internal estimates since they enable the understanding of different factors which enter into the variance in order to guide experimental design and data analysis.  However Jackknife estimates are less sensitive to failures in distributional assumptions, and this provides them with statistical robustness. 

It is conventional to display spectra on a log scale.  This is because taking the log of the spectrum stabilizes the variance and leads a distribution which is approximately Gaussian.   
\subsection{Confidence intervals}
\label{conf1}

The expected values of the estimators and also their variance have been discussed for several different spectral estimators but it is desirable to put confidence intervals on the spectral estimates rather than standard deviations.

As previously mentioned in section \ref{trials} the averaging of direct spectral estimates from different trials yields, in the large sample limit, estimates which are distributed as $\chi_{2N_T}^{2}$.  In general for the other estimates a well known approximation \cite{pw1} is to assume that the estimate is distributed as $\chi_{\nu_0}^{2}$.  Confidence intervals can then by placed on estimates on the basis of this $\chi_{\nu_0}^{2}$ distribution.  The confidence interval applies for the population spectrum $S(f)$ and is obtained from the following argument. 

\begin{equation}
P\left[q_1 \le \chi_{\nu_0}^{2} \le q_2\right] = 1 - 2p
\end{equation}
Where $P$ indicates probability, $q_1$ is such that $P[\chi_{\nu_0}^{2} \leq q_1] = p$ and $q_2$ is such that $P[\chi_{\nu_0}^{2} \geq q_2] = p$.  It follows that,

\begin{equation}
P\left[q_1 \le \nu_0 I^{X}(f) / S(f) \le q_2\right] = 1 - 2p
\end{equation}
Hence an approximate $100\% \times (1-2p)$ confidence interval for $S(f)$ is given by,

\begin{equation}
P\left[\nu_0 I^{X}(f)/q_2 \le S(f) \le \nu_0 I^{X}(f)/q_1\right]
\label{conf2} 
\end{equation}

For large $\nu_0$ ($>$ 20) these confidence intervals do not differ substantially from those based on a Gaussian ($\pm$2 standard deviations) but at small $\nu_0$ the difference can be substantial as for these values the $\chi_{\nu_0}^{2}$ distribution has long tails.

\subsection{High Frequency limit}

The population spectrum goes to a constant value equal to the rate $\lambda$ in the high frequency limit.  In practice spectra calculated from a finite sample will go to a value close to $\lambda$ but fluctuations in the number of spikes in the interval will lead to an error in this estimate.  For a given sample the spectrum will go the value given by equation \ref{HFL}. 

\begin{equation}
I(f \rightarrow \infty) = \frac{1}{N_TK} \sum_{k=0}^{K-1} \sum_{n=1}^{N_T} \sum_{j=1}^{N_n(T)} h_k(t^n_j)^2 \label{HFL}
\end{equation}
Where $t^n_j$ is the $j^{th}$ spike in the $n^{th}$ trial and $N_n(T)$ is the total number of spikes in the $n^{th}$ trial.  In the case of direct and lag window estimators the averaging over tapers need not be performed.\\

This expression yields a value which is typically very close to the sample estimate of the mean rate\footnote{It is exactly the sample estimate of the mean rate for a rectangular taper.}.  It is significant departures from this high frequency limit which are of interest when interpreting the spectrum as these indicate enhancement or suppression relative to a homogeneous Poisson process.

\subsection{Choice of estimator, taper and lag window}

The preceding section discussed the large sample statistical properties of direct, lag window and multitaper estimates of the spectrum.  The choice of which estimator to use remains a contentious one \cite{pw1}.  The multitaper method is the most systematic of the estimators but the lag window estimators should perform almost as well for those spike train spectra which have reasonably small dynamic ranges\footnote{Dynamic range is a measure of the variation in the spectrum as a function of frequency and is defined as $10 log_{10} (\frac{max_f S(f)}{min_f S(f)})$.}.  However, it is possible to construct spike trains with widely different time scales, which can possess a large dynamic range.  In addition, the multitaper technique leads to a simple jackknife procedure by leaving out one data taper in turn.  A further important property of the multitaper estimator is that it gives more weight to events at the edges of the time interval and thus ameliorates the arbitrary downweighting of the edges of the data introduced by single tapers.

 If using the lag window estimator there are many choices available for both the taper and the lag window.  The choice of taper is generally not critical provided that the taper goes smoothly to zero at the start and end of the interval.  A rectangular taper has particularly large sidelobes in the frequency domain which can lead to significant bias.  The choice of lag window is also usually not critical and typically a Gaussian kernel will be satisfactory. 

\section{Estimating the Coherency}
\label{estthecoh}

Sample coherency, which may be estimated using equation \ref{Coh0}, may be evaluated using any of the previously mentioned spectral estimators.  The principle difference is that the direct estimator, in terms of which the other estimators are expressed, is given by equations \ref{Coh1} and \ref{Coh2} rather than \ref{D1} and \ref{D2}.
\hspace{-3.5mm}
\begin{eqnarray}
C(f) & = & I^X_{12}/\sqrt{I^X_{11}I^X_{22}} \label{Coh0} \\
I^{D}_{12}(f) & = & J^D_1(f)J^D_2(f)^* \label{Coh1} \\
J^{D}_a(f) & = & \int_{-\infty}^{\infty} h(t) e^{-2 \pi i f t} d\overline{N}_a(t) \label{Coh2}
\end{eqnarray}
Where the $\overline{N}_1(t)$ and $\overline{N}_2(t)$ are simultaneously recorded spike trains from two different cells and $X$ denotes the type of spectral estimator.  Possible choices of estimator $X$ include; $D$ direct, $DT$ trial averaged direct, $LW$ lag window or $MT$ multitaper.\\

Lag window and multitaper coherency estimates may be constructed by substituting $I^{D}_{12}(\cdot)$ in place of $I^{D}(\cdot)$ in equations \ref{LW1} and \ref{MT1}.  The estimates are biased over a frequency range equal to the width of the smoothing although the exact form for the bias is difficult to evaluate.

\subsection{Confidence limits for the Coherence}
\label{cohconf}

The treatment of error bars is somewhat different between the spectrum and the coherency, since the coherency is a complex quantity.  Usually one is interested is in establishing whether there is significant coherence in a given frequency band.  In order to do this the sample coherence should be tested against the null hypothesis of zero population coherence.  The distribution of the sample coherence under this null hypothesis is given below.  

\begin{equation}
P(|C|) = (\nu_0 - 2)|C|(1-|C|^2)^{\hspace{0.5mm} (\nu_0/2-2)} \hspace{2cm} 0 \le |C| \le 1 \label{Nsph}  
\end{equation}

A derivation of this result is given in \cite{Hannan}.  In outline the method is to rewrite the coherence in such a way that it is equivalent to a multiple correlation coefficient \cite{Anderson}.  The distribution of a multiple correlation coeffient is then a known result from multivariate statistics.   In the case of coherence estimates based on lag window estimators the appropriate $\nu_0$ may be used although this is only approximately valid because this method of derivation assumes integer $\nu_0/2$. 

It is straightforward to calculate a confidence level based on this distribution.  The coherence will only exceed $\sqrt{1-p^{1/(\nu_0/2-1)}}$ in $p\times 100\%$ of experiments.  In addition it is notable that the quantity $(\nu_0/2-1)|C|^2/(1-|C|^2)$ is distributed as $F_{2,\nu_0-2)}$ under this null hypothesis.  It is useful to apply a transformation to the coherence before plotting it which aids in the assessment of significance.  The variable $q = \sqrt{-(\nu_0 -2)log(1-|C|^2)}$ has a Raleigh distribution which has density $p(q) = qe^{-q^2/2}$.  This density function does not depend on $\nu_0$ and furthermore has a tail which closely resembles a Gaussian.  For certain values of a fitting parameter\footnote{A reasonable choice for $\beta$ is 23/20.} $\beta$, a further linear transformation $r = \beta (q-\beta)$ leads to a distribution which closely resembles a standard normal Gaussian for $r > 2$.  This means that for $r > 2$ one can interpret $r$ as the number of standard deviations by which the coherence exceeds that expected under the null hypothesis.  

\subsection{Confidence Limits for the Phase of the Coherency} 

If there is no population coherency then the phase of the sample coherency is distributed uniformly.  If, however, there is population coherency then the distribution of the sample phase is approximately Gaussian provided that the tails of the Gaussian do not extend beyond a width $2 \pi $.  An approximate 95\% confidence interval for the phase \cite{rosen,brts} is given below.

\begin{equation}
\hat{\phi}(f) \pm 2 \sqrt{\frac{2}{\nu_0} \left( \frac{1}{|C(f)|^2} - 1 \right) \label{cphsc}}
\end{equation}
Where $\hat{\phi}(f)$, the sample estimate of the coherency phase, is evaluated using \linebreak $tan^{-1}\{Im(C)/Re(C)\}$. 

\section{Finite Size Effects}
\label{fse}

In the preceding sections error bars were given for estimators of the spectrum and the coherence.  However these error bars were based on large sample sizes (they apply asymptotically as $T \rightarrow \infty$).  Neurophysiological data are not collected in this regime and, particularly in awake behaving studies where data is often sparse, corrections arising at small T are potentially important.  In order to estimate the size of these corrections a particular model for the point process is required.  The model studied was chosen primarily for its analytical tractability while still maintaining a non-trivial spectrum.  

\begin{figure}[hbtp] 
\begin{center}
\leavevmode
\hbox{%
\unitlength=1mm
\begin{picture}(0,32)
\put(-70,0){\includegraphics[height=2.25cm]{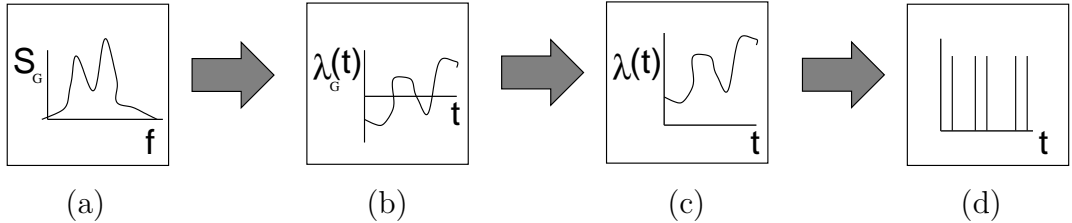}}
\put(-62,-6){(a)}
\put(-22,-6){(b)}
\put(18, -6){(c)}
\put(57, -6){(d)}
\end{picture}}
\end{center}

\caption{Schematic illustrating the model for which finite size corrections to the asymptotic error bars will be evaluated. (a) A spectrum $S_G(f)$ is defined (b) A realization $\lambda_G(t)$ is drawn from a Gaussian process with this spectrum (c) The mean rate $\lambda$ is added to $\lambda_G(t)$ to obtain $\lambda(t)$ (d) This rate function is used to generate a realization of an inhomogeneous Poisson process yielding a set of spike times.}
\label{dsspic}
\end{figure}

The model and the final results will be presented here but the details of the analysis are reserved until appendix \ref{app1}.  The model is a doubly stochastic inhomogeneous Poisson process with a Gaussian rate function.  A specific realization of a spike train is generated from the model in the following manner.  Firstly a population spectrum $S_G(f)$ is specified.  From this a realization of a zero mean Gaussian process $\lambda_G(f)$ is generated.  A constant $\lambda$, the mean rate, is then added to this realization.   This function is then considered to be the rate function for an inhomogeneous Poisson process.  A realization of this inhomogeneous Poisson process is then generated.  A schematic of the model is shown in figure \ref{dsspic}.

Technically this is not a valid process because the rate function $\lambda(t)$ may be negative.  However, if the area underneath the spectrum is small enough then the fluctuations about the mean rate seldom cross zero and corrections due to this effect are negligible.  In addition large violations of this area constraint have been tested by Monte Carlo simulation and the results still apply to a good approximation.

An important feature of this model is that the population spectrum of the spike trains is simply the spectrum of the inhomogeneous Poisson process rate function plus an offset equal to the mean rate\footnote{This result does not depend on the Gaussian assumption.}.  The spectrum of the rate function is a positive real quantity and therefore within this model the population spectrum cannot be less than the mean rate at any frequency.  Intuitively, the reason for this is that the process must be more variable than a homogeneous Poisson process at all frequencies.

To make the nature of the result clear a simplified version is given in equation \ref{mainres1}.  This version is for the particular case of a homogeneous Poisson process (which has a flat population spectrum) and a rectangular taper\footnote{The expression also holds approximately for the multitaper estimate provided all tapers up to the Shannon limit are used.}.

\begin{equation}
var\{I^X(f)\} =  \lambda^2 \left [\frac{2}{\nu_0} + \frac{1}{N_T T \lambda} \right ]
\label{mainres1}
\end{equation}
Where $\lambda$ is the mean rate.\\

A sample based estimate of $N_T T \lambda$ is the total number of spikes over all trials.  It is to be noted that finite size effects reduce the degrees of freedom.  This result implies that there is a point beyond which additional smoothing does not decrease the variance further and this point is approximately when $\nu_0$ is equal to twice the total number of spikes.  The full result is given in equations \ref{mainres2} - \ref{mainres6}.  

\begin{eqnarray}
& & var\{I^X(f)\}  =  \frac{2 E\{I^X(f) \}^2}{\nu(f)} \label{mainres2}\\
& & \frac{1}{\nu(f)} =  \frac{1}{\nu_0} + \frac{C^X_h\Phi(f)}{2TN_TE\{I^X(f) \}^2} \label{mainres2.5}
\end{eqnarray}
Where,
\begin{equation}
C^X_h = 
\left \{ \begin{array}{l} 
\int_0^1 f(t)^4 dt \hspace{5.05cm} \rm{If} \hspace{5mm} X = \rm{LW,D \hspace{2mm} or \hspace{2mm} DT}\\
\frac{1}{K^2} \sum\limits_{k,k'} \int_0^1 f_k(t)^2 f_{k'}(t)^2 dt \hspace{2.65cm} \rm{If} \hspace{5mm} X = \rm{MT}   
\end{array} \right . \label{mainres3}
\end{equation}
\begin{equation}
f(t/T) = \sqrt{T} h(t) \label{mainres4}
\end{equation}
and,
\begin{equation}
\Phi(f) = \lambda_{hf} + 4[E\{I^X(f)\} - \lambda_{hf}] + 2[E\{I^X(0)\} - \lambda_{hf}] + [E\{I^X(2f)\} - \lambda_{hf}]    \label{mainres5} 
\end{equation}
\begin{equation}
\lambda_{hf}  = E \{ I^X(f \rightarrow \infty) \} \label{mainres6}
\end{equation}

$C_h^X$ is a constant of order unity which depends on the taper.  When a taper is used to control bias some of the spikes are effectively disregarded and this has an effect on the size of the correction.  The function $f(t)$ has the same form as the taper $h(t)$ but is defined for the interval $[0,1]$.  $C_h^X$ is the integral of the fourth power of $f$ and obtains its minimum value of one for a rectangular taper.  It is typically between 1 and 2 for other tapers.  In the multitaper case cross terms between tapers are included.\\

Equation \ref{mainres5} describes how the finite size correction depends on the structure of the spectrum.  $\Phi(f)$ is the sum of four terms.  The first term is the only term which is present for a flat spectrum.  The second term is a correction which depends on the spectrum at the frequency being considered.  The next two terms depend on the spectrum at zero frequency and the spectrum at twice the frequency being considered.  The latter three terms all depend on the difference between the spectrum at some frequency and the high frequency limit.  Equation \ref{mainres5} applies provided that the spike train is well described by the model.  However, this is not necessarily the case and a suppression of the spectrum, which cannot be described by the model, often occurs at low frequencies\footnote{Note that any spike train spectra displaying significant suppression below the mean firing rate can immediately rule out the inhomogeneous Poisson process model.}.  In the event that there is a significant suppression of the spectrum $\Phi(f)$ may become small or even negative.  To avoid this a modified form for $\Phi(f)$ may be used which prevents this.   
\begin{eqnarray}
\Phi(f) = \lambda_{hf} + 4\rm{max}([E\{I^X(f)\} - \lambda_{hf}],0) & + & 2\rm{max}([E\{I^X(0)\} - \lambda_{hf}],0) ... \nonumber \\
   & + & \rm{max}([E\{I^X(2f)\} - \lambda_{hf}],0)    \label{mainres7}
\end{eqnarray}

The above modification to the result is somewhat {\it ad hoc} so Monte Carlo simulations of spike trains with enforced refractory periods have been performed to test its validity.  These simulations demonstrated that, although the correction derived using \ref{mainres7} was significantly different from that obtained from the Monte Carlo simulations in the region of the suppression, equation \ref{mainres7} provided a pessimistic estimate in all cases studied.  This increases confidence that applying finite size corrections using equation \ref{mainres7} will provide reasonable error bars for small samples.  

Equation \ref{mainres2.5} gives the finite size correction in terms of a reduction in $\nu_0$.  The new $\nu(f)$ may be used to put confidence intervals on the results, as described in section \ref{conf1}, although the accuracy of the $\chi_{\nu}^{2}$ assumption will be reduced.  In the case of the coherence an indication of the correction to the confidence level can be obtained by using the smaller of the two $\nu(f)$ from the spike train spectra to calculate the confidence level using equation \ref{Nsph}.  In all cases if the effect being observed only achieves significance by an amount which is of the same order as the finite size correction then it is recommended that more data be collected.

\section{Experimental Design}

Often it is useful to know in advance how many trials or how long a time interval one needs in order to resolve features of a certain size in the spectrum or the coherence.  To do this one needs to estimate the asymptotic degrees of freedom $\nu_0$.  This depends on the size of feature to be resolved $\alpha$, the significance level for which confidence intervals will be calculated $p$ and the fraction of experiments which will achieve significance ${\cal P}$. In addition the reduction in the degrees of freedom due to finite size effect depends on the total number of spikes $N_s$ and also $C_h$ (see section \ref{fse}).

An estimate of $v_0$ may be obtained in two stages.  Firstly $\alpha$,$p$ and ${\cal P}$ are specified and used to calculate a degrees of freedom $\nu$.  Secondly the asymptotic degrees of freedom $\nu_0$ is estimated using $\nu$, $N_s$ and $C_h$.  The feature size $\alpha = (S-\lambda)/\lambda$ is the minimum size of feature which the experimenter is content to resolve.  For example, a value of 0.5 indicates that where the population spectrum exceeds $1.5 \lambda$ the feature will be resolved.  The significance level should be set to the same value that will be used for calculating the confidence interval for the spectrum, typically be $0.05$.  For a given $p$ there is some probability ${\cal P}$ that an experiment will achieve significance.  To calculate $\nu$ one begins with a guess $\nu_g$.  Then $q_1$ is chosen such that $P\left[\chi^2_{\nu_g} \ge q_1\right] = p/2$.  On the basis of this one then evaluates ${\cal P_0} = 1 - \Phi \left[ q_1/(1 + \alpha) \right]$ where $\Phi$ is the cumulative $\chi^2_{\nu_g}$ distribution\footnote{These formulae apply for $\alpha > 0$.  If $\alpha < 0$ then $P\left[\chi^2_{\nu_g} \le q_1\right] = p/2$ and ${\cal P_0} = \Phi \left[ q_1/(1 + \alpha) \right]$ should be used.}.  If ${\cal P_0}$ is equal to the specified fraction ${\cal P}$ then $\nu = \nu_g$ otherwise a different $\nu_g$ is chosen.  This procedure is readily implemented as a minimization of $({\cal P - P_0}(\nu_g))^2$ on a computer.  Having obtained $\nu$ one can estimate $\nu_0$ using the following expression.

\begin{equation}
\frac{1}{\nu_0} = \frac{1}{\nu} - \frac{C_h \left[ 1 + 4 \alpha \right]}{2 N_s \left[ 1 + \alpha \right]^2} 
\end{equation}
Where the $4\alpha$ is omitted from the numerator if $\alpha < 0$.

\begin{figure}[hbtp] 
\begin{center}
\leavevmode
\hbox{%
\unitlength=1mm
\begin{picture}(0,50)
\put(-35,-5){\includegraphics[height=5cm]{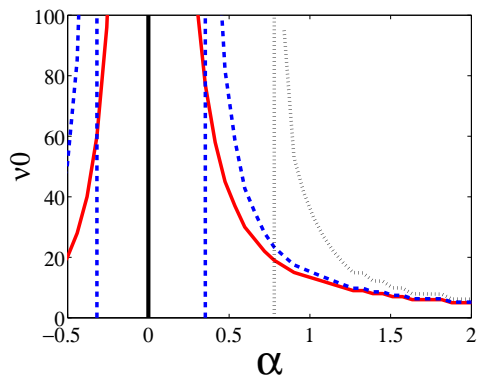}}
\end{picture}}
\end{center}
\caption{Example design curves for the case when $p = 0.05$, ${\cal P} = 0.5$ and $C_h = 1.5$.  The three curves correspond to $N_s = \infty$ (solid), $N_s = 100$ (dashed), $N_s = 50$ (dotted).}
\label{expdes}
\end{figure}

Figure \ref{expdes} illustrates example design curves generated using this method.  These curves show the asymptotic degrees of freedom as a function of feature size for different total numbers of spikes.

The existence of a region bounded by vertical asymptotes implies that as long as the total number of measured spikes is finite, modulations in the spectrum below a certain level cannot be detected no matter how much the spectrum is smoothed.  These curves may be used to design experiments capable of resolving spectral features of a certain size.

In the case of the coherence one calculates how many degrees of freedom are required for the confidence line to lie at a certain level as described in section \ref{cohconf}.

\section{Line Spectra}
\label{linspec1}

One of the assumptions underlying the estimation of spectra is that the population spectrum varies slowly over the smoothing width ($W$ for multitaper estimators).  While this is often the case there are situations in which the spectrum contains very sharp features which are better approximated by lines than by a continuous spectrum.  This corresponds to periodic modulations of the underlying rate, such as when a periodic stimulus train is presented.  In such situations it is useful to be able to test for the presence of a line in a background of colored noise (i.e. in a locally smooth but otherwise arbitrary continuous population spectrum).  Such a test has been previously developed, in the context of multitaper estimation, for continuous processes \cite{dt1} and in the following section the analogous development for point processes is presented.  

\subsection{F-test for point processes}  
\label{ftestsec}

A line in the spectrum has an exactly defined frequency and consequently the process $N(t)$ has a non-zero first moment.  The natural model in the case of a single line is given by equation \ref{FT1}.

\begin{equation}
E\{dN(t)\}/dt = \lambda_0 + \lambda_1 cos(2 \pi f_1 t + \phi) \label{FT1} \\
\end{equation}

A zero mean process ($\overline{N}$) may be constructed by subtraction of an estimate of $\lambda_0t$.  Provided that the product of the line frequency($f_1$) and the sample duration(T) is much greater than one the sample quantity $N(T)/T$ is an approximately unbiased estimate of $\lambda_0$.  The resultant zero mean process $\overline{N}$ has a Fourier transform which has a non-zero expectation.

\begin{eqnarray}
J_k(f) & = & \int_{-\infty}^{\infty} h_k(t) e^{-2\pi i f t} d\overline{N}(t) \\
E\{ J_k(f)\} & = & c_1 H_k(f-f_1) + c_1^* H_k(f+f_1)
\end{eqnarray}
Where,
\begin{equation}
c_1  =  \lambda_1 e^{i\phi} /2 
\end{equation}
In the case where $f > 0$ and $f_1 > W$,
\begin{equation}
E\{ J_k(f)\}  \simeq  c_1 H_k(f-f_1)
\end{equation}

The estimates of $J_k(f_1)$ from different tapers provide a set of uncorrelated estimates of $c_1 H_k(0)$.  It is hence possible to estimate the value of $c_1$ by complex regression.

\begin{equation}
\hat c_1 = \frac{\sum_k J_k(f_1)H_k(0)}{\sum_k | H_k(0)|^2} \label{estc1}
\end{equation}

Under the null hypothesis that there is no line in the spectrum ($c_1 = 0$) it may readily be shown that $E\{ \hat c_1 \} = 0$ and $var\{\hat c_1 \} = S(f_1)/ \sum_k | H_k(0)|^2$.  The residual spectrum\footnote{It is also possible to estimate a residual coherency.  In order to do this one uses a residual cross-spectrum $\hat S_{xy}(f) = \frac{1}{K} \sum_k ( J_k^x(f) - \hat c_1^x H_k(f-f_1))^*( J_k^y(f) - \hat c_1^y H_k(f-f_1))$, together with the residual spectra to evaluate the usual expression for coherency.}, which has the line removed, may be estimated using equation \ref{resspec}.

\begin{equation}
\hat S(f) = \frac{1}{K} \sum_k | J_k(f) - \hat c_1 H_k(f-f_1)|^2 \label{resspec}
\end{equation}

In the large sample limit the distributions of both $\hat c_1$ and $\hat S(f_1)$ are known \cite{pw1} and may be used to derive relation \ref{fdist}. 

\begin{equation}
\frac{|\hat c_1 |^2 \sum_k |H_k(0)|^2(K-1)}{\sum_k |J_k(f_1) - \hat c_1 H_k(0)|^2} \doteq F_{2,2(K-1)} \label{fdist}
\end{equation}
Where $\doteq$ denotes `is distributed as'.
 
The null hypothesis may be tested using this relation and, if rejected, the line can be removed using equation \ref{resspec} to estimate the residual spectrum.  It is worth noting that although relation \ref{fdist} was derived for large samples the test is remarkably robust as the sample size is decreased.  Numerical tests indicate that the tail of the F distribution is well reproduced even in situations where there are as low as 5 spikes in total.

\subsection{Periodic Stimulation}

A common paradigm in neurobiology where line spectra are particularly important is that of periodic stimulation.  When a neuron is driven by a periodic stimulation of frequency $f_1$ the spectrum may contain lines at any of the harmonics $nf_1$.  Provided that $f_1 > 2W$ the analysis of section \ref{ftestsec} applies with each harmonic being separately tested for significance.  

The first moment of the process, which has period $1/f_1$, is given by equation \ref{1stmom} and may be estimated using $\hat c_n$.

\begin{equation}
\lambda(t) = \lambda_0 + \sum_n \lambda_n cos(2 \pi n f_1 t + \phi_n) \label{1stmom}
\end{equation}
Where $\lambda_n = 2 |c_n|$, $\phi_n = \tan^{-1} \{ Im(c_n)/Re(c_n) \}$, the sum is taken over all the significant coefficients.

This rate function $\lambda(t)$ is the average response to a single stimulus or impulse response.  The coefficients $c_n$ are the Fourier series representation of $\lambda(t)$.

\vspace{-1mm}
\subsection{Error Bars}

It is possible to put confidence intervals on both the modulus and the phase of the coefficients $\hat c_n$.   For large samples($> 10$ spikes) the real and imaginary parts of $\hat c_n$ are distributed as independent Gaussians each with standard deviation $\sigma_n = \sqrt{S(nf_1)/(2 \sum_k |H_k(0)|^2)}$.  For $c_n/\sigma_n >3$ the distribution of $|\hat c_n|$ is well approximated by a Gaussian centered on $|c_n|$ and with standard deviation $\sigma_n$.  In addition the estimated phase angle ($\hat \phi_n$) is also almost Gaussian with mean $\phi_n$ and standard deviation $\sigma_n /|c_n|$.  Approximate error bars or confidence intervals may be obtained using a sample based estimate of $\sigma_n$, $\hat \sigma_n = \sqrt{\hat S(nf_1)/(2 \sum_k |H_k(0)|^2)}$.

Estimating error bars for the impulse response function is more involved due to their non\--local nature (if one of the Fourier coefficients is varied the impulse response function changes everywhere).  It is therefore of interest to estimate a global confidence interval, defined as any interval such that the  probability of the function crossing the interval {\it anywhere} is some predefined probability.  A method for estimating a global confidence band is detailed in \cite{Clive1} and outlined here.  First a basis vector $\Phi(t)$ is constructed.

\begin{equation}
\Phi(t) =  \left[ \begin{array}{c}
           \hat \sigma_1 cos (2 \pi f_1 t) \\
            \vdots \\
           \hat \sigma_N cos (2 \pi f_N t) \\
           \hat \sigma_1 sin (2 \pi f_1 t) \\
            \vdots \\
           \hat \sigma_N sin (2 \pi f_N t)
           \end{array}
           \right]
\end{equation}
Where $N$ is the total number of harmonics. 

The elements of this vector have unit variance and a standard approximation may be applied.

\begin{equation}
P(sup | \lambda(t) - E\{\lambda(t)\} | > c || \Phi(t) || ) \le 2(1-N(c)) + (k/\pi) e^{-c^2/2}
\end{equation}
Where $sup$ is the maximum value of its operand, $||\Phi(t)||$ denotes the length of vector $\Phi(t)$, $N(c)$ is the cumulative standard normal distribution and $k$ is a constant.  $k$ may be evaluated by constructing the $2\times N$ matrix $X(t) = [\Phi(t) \hspace{1mm} d\Phi(t)/dt ] $, forming its $QR$ decomposition \cite{numrec} and then evaluating $k = \int_0^T |R_{22}(t)/R_{11}(t)| dt$. 

Confidence intervals for the residual spectrum are calculated in the usual manner (using $\chi_{\nu}^2$) although at the line frequencies the interval is slightly broadened due to the loss of a degree of freedom incurred by estimation of $c_n$.  Section \ref{periodiceg} contains an example application of the methods described in this section.

\section{Example Spectra}

\begin{figure}[hbtp] 
\begin{center}
\leavevmode
\hbox{%
\unitlength=1mm
\begin{picture}(0,50)
\put(-73,0){\includegraphics[height=3.8cm]{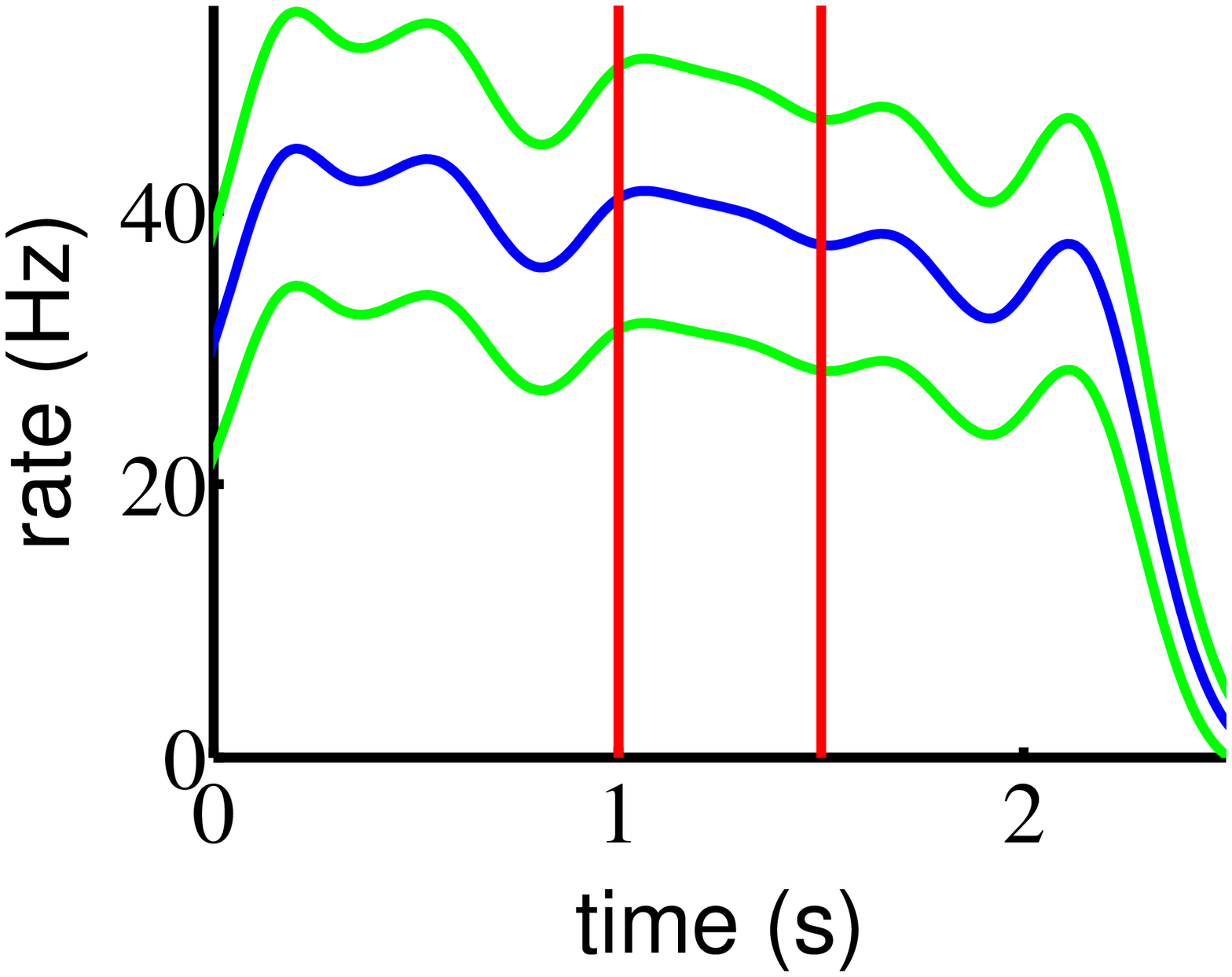}}
\put(-28,0){\includegraphics[height=3.8cm]{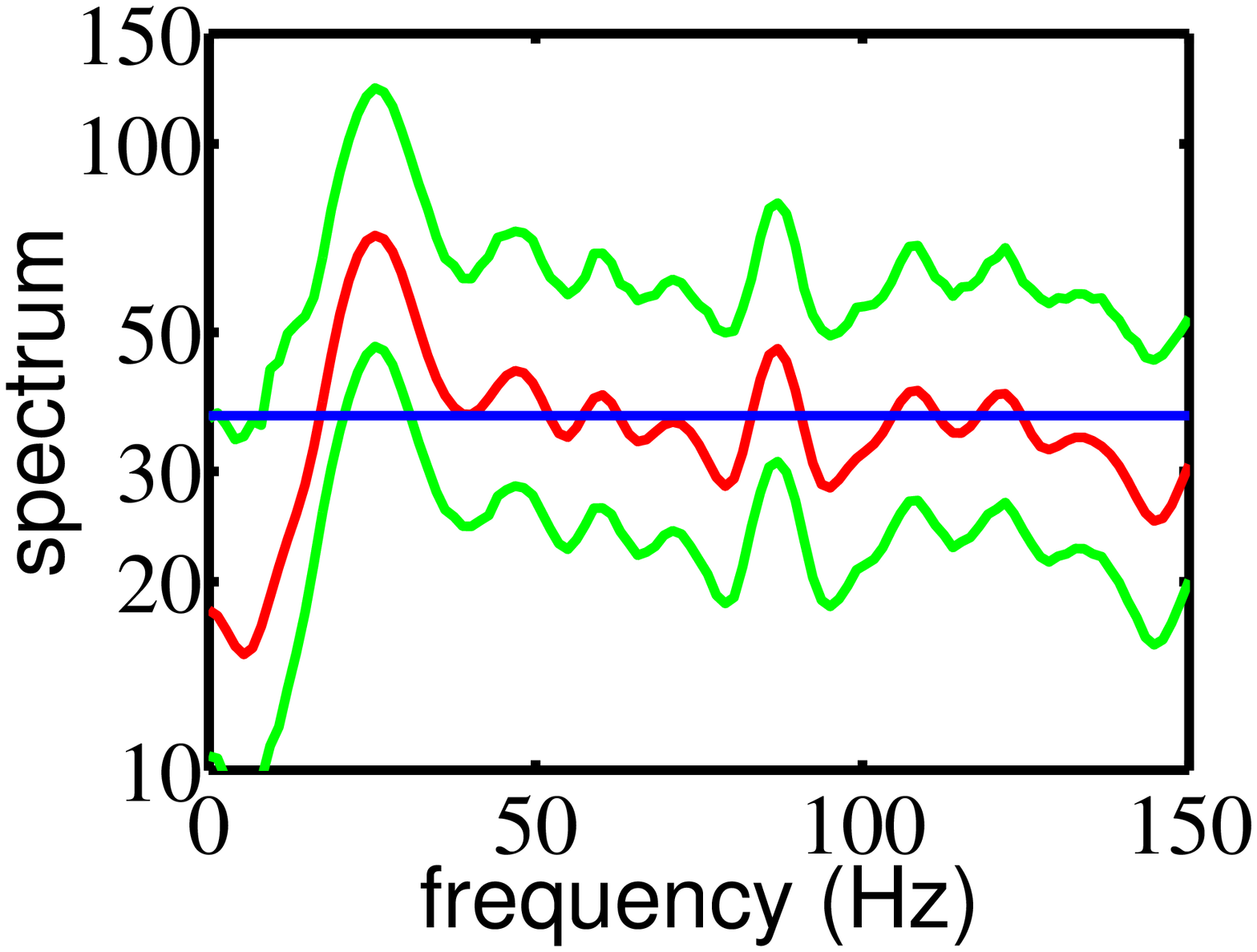}}
\put(22,0){\includegraphics[height=3.8cm]{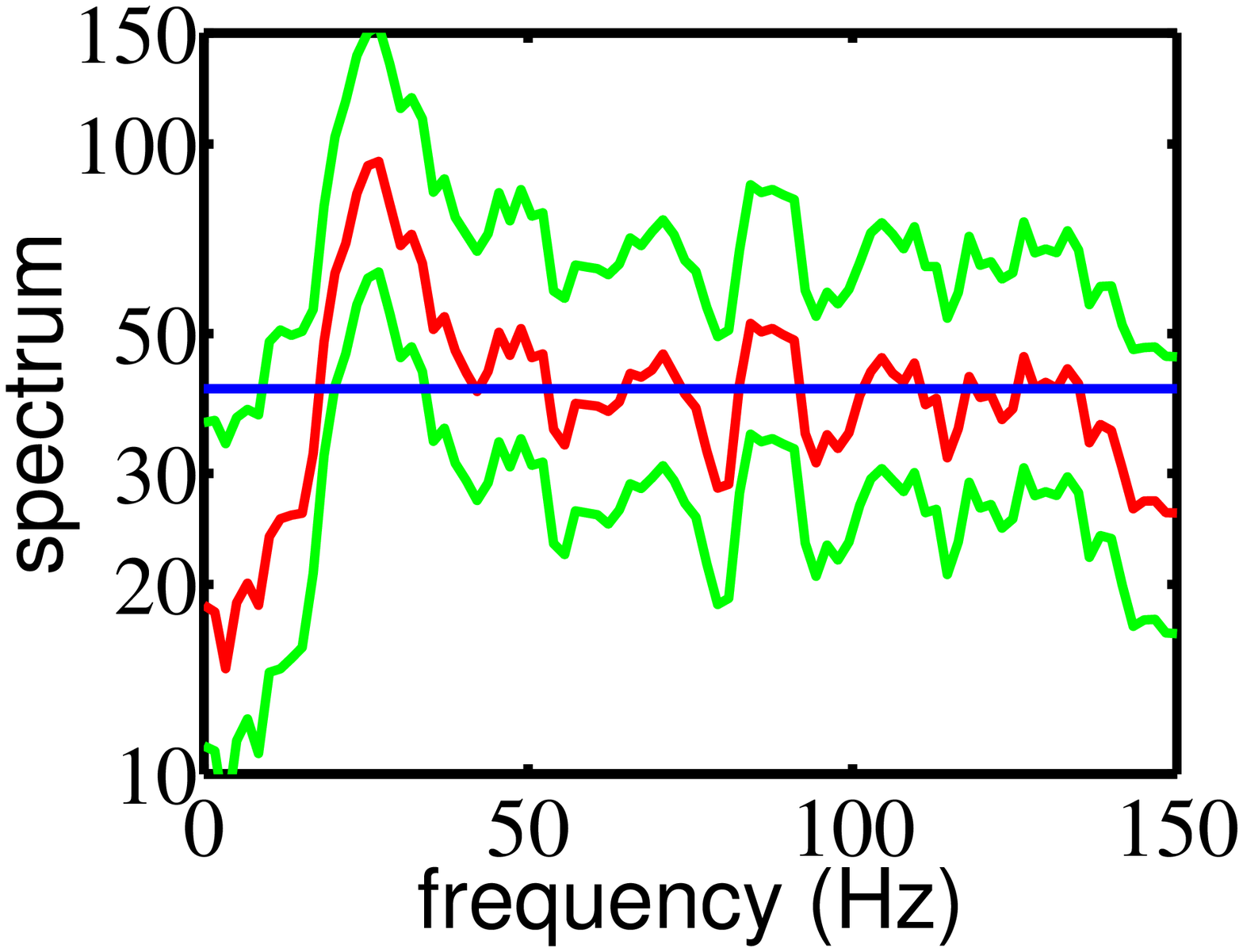}}
\put(-50,-5){(a)}
\put(0,-5){(b)}
\put(50, -5){(c)}
\end{picture}}
\end{center}
\caption{(a) Gaussian kernel (100 ms width) smoothed firing rate with $2\sigma$ error bars based on a stationarity assumption. The vertical lines indicate the period over which the spectrum was calculated.  A light is flashed at time zero and the spectrum is evaluated over the interval when the monkey is required to remember the target location. (b) The spectrum evaluated over this interval using a lag window estimator with a 40\% cosine taper and a Gaussian lag window of width 3.5 Hz.  95\% confidence limits are shown with the finite size correction included (this typically resulted in a decrease in $\nu(f)$ from about 50 to 36).  The horizontal line indicates the high frequency limit.(c) The same spectrum evaluated using a multitaper estimator.  A bandwidth (W) of 5 Hz was used allowing 5 tapers.  Both estimators have the same degrees of freedom.}
\label{spectrum1}
\end{figure}

Figure \ref{spectrum1} is a spectrum calculated from data collected from a single cell recorded from area PRR in the parietal cortex of an awake behaving monkey during a delayed memory reach task \cite{Reach1}.  The spectrum is calculated over an interval of 0.5s during which the firing rate is reasonably stationary and is averaged over 5 trials.  The spectrum shows two features which achieve significance.  There is enhancement of the spectrum in the frequency band 20-40 Hz indicating the presence of an underlying broad band oscillatory mode in the neuronal firing rate.  In addition there is suppression of the spectrum at low frequencies.  As discussed previously a suppression of this sort is consistent with an effective refactory period during which the neuron is less likely to fire.  Care must be taken at low frequencies since at frequencies comparable to the smoothing width the spectrum is particularly sensitive to any non-stationarity in the data. 

\vspace{-2mm}
\section{Example Coherency}

\begin{figure}[hbtp] 
\begin{center}
\leavevmode
\hbox{%
\unitlength=1mm
\begin{picture}(0,50)
\put(-70,0){\includegraphics[height=5cm]{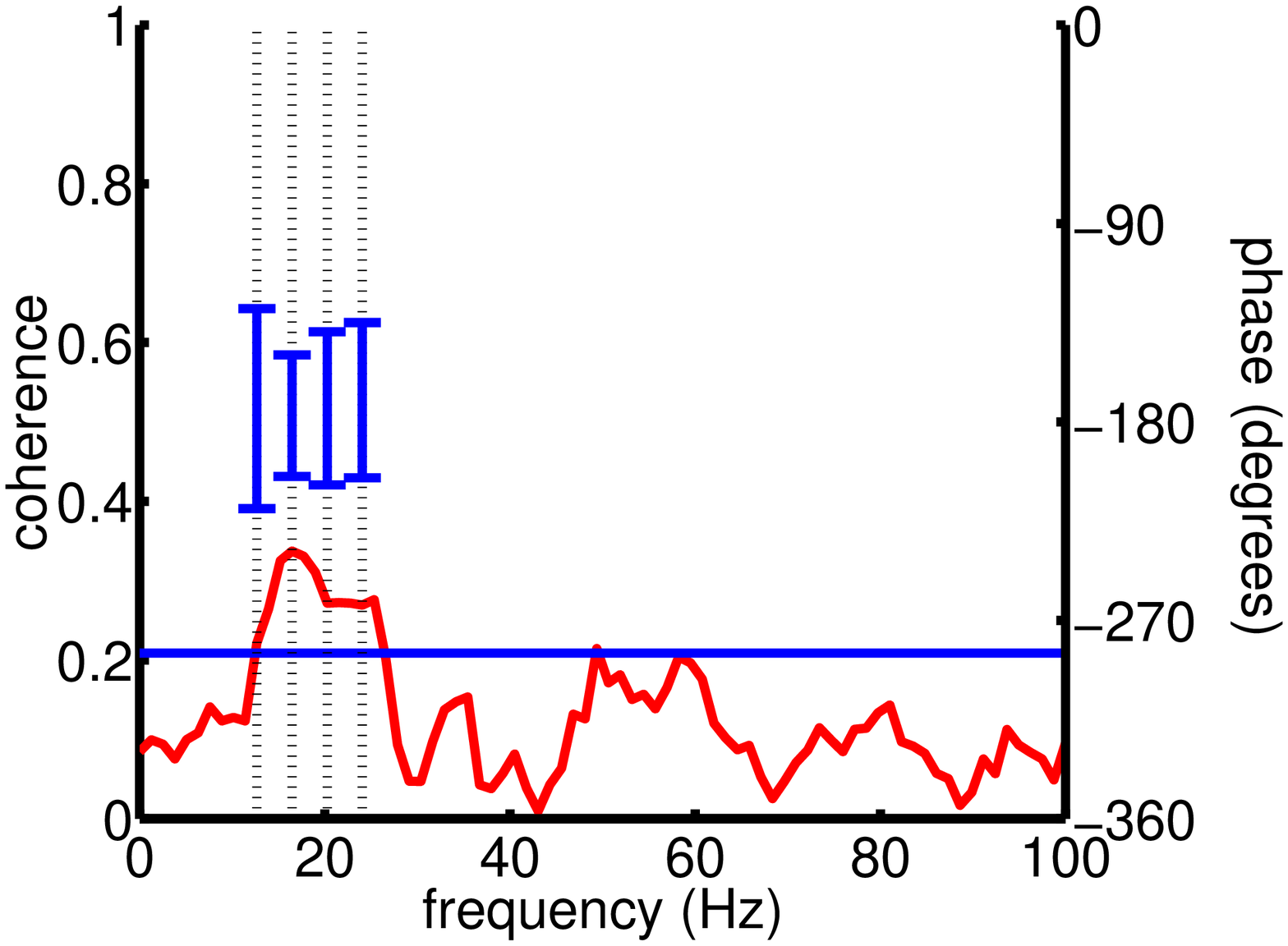}}
\put( 5,0){\includegraphics[height=5cm]{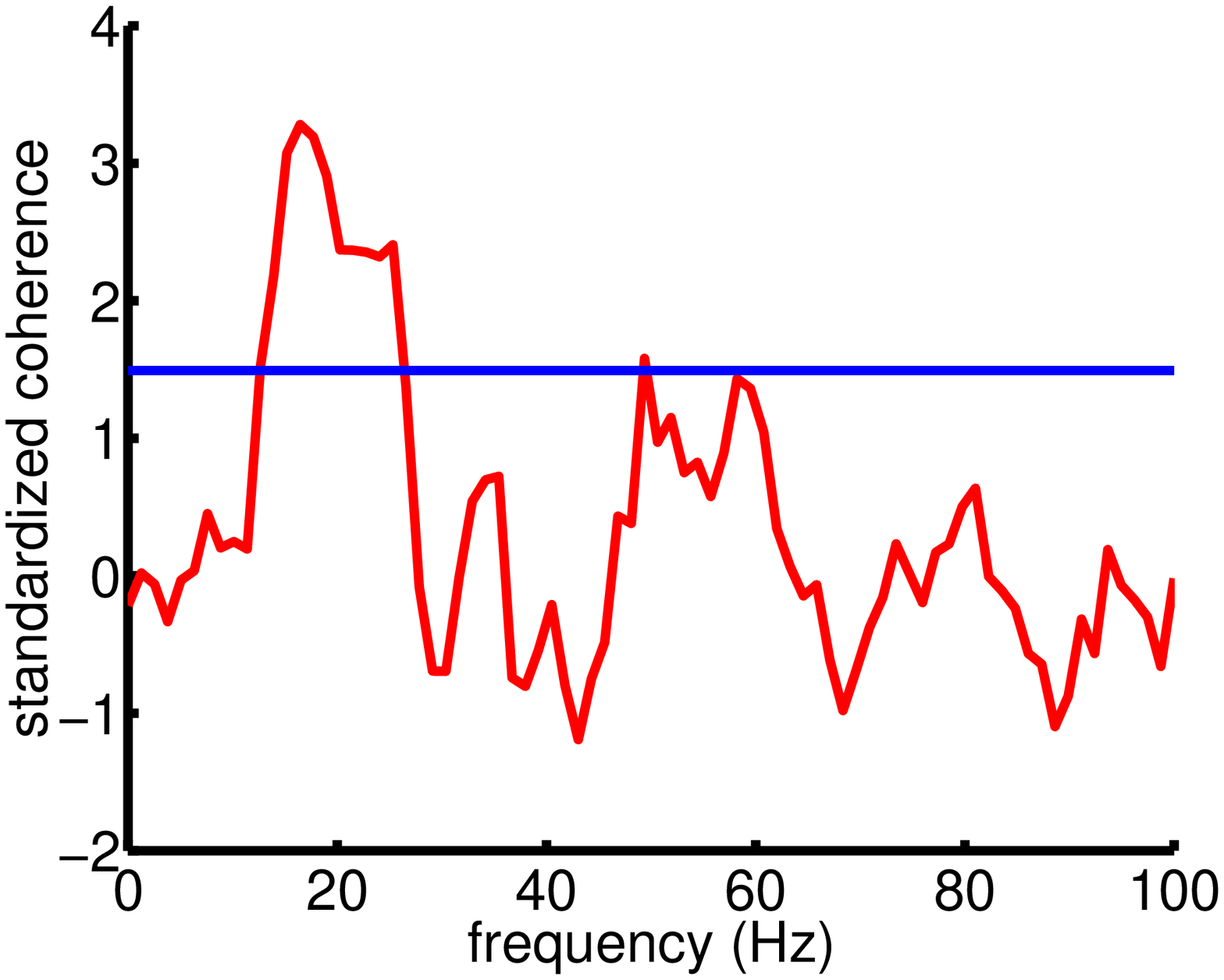}}
\put(-40,-5){(a)}
\put(35, -5){(b)}
\end{picture}}
\end{center}
\caption{(a) Coherence (left axis) and the phase of the coherency (right axis).  Fifteen trials of 0.5s duration were simulated using a doubly stochastic Poisson process as described in the text.  A multitaper estimator with a smoothing width of 7 Hz was used.  Finite size corrections were used and resulted in 25\% reduction in the degrees of freedom. A horizontal line has been drawn at the 95\% confidence level under the null hypothesis of no coherency.  Where the null hypothesis is rejected the phase of the coherency is estimated and shown with an approximate 95\% confidence interval.  (b) The standardized coherence is a transformation which maps the null distribution onto an approximately standard normal variate (as described in section \ref{cohconf}).  The estimated coherence at 20 Hz would therefore lie at three standard deviations if there were no population coherence.}
\label{coherency1}
\end{figure}

To illustrate the estimation of coherency simulated spike trains were generated from a coupled doubly stochastic Poisson process.  For a given trial a pair of rate functions were drawn from a Gaussian process.  The realizations share a coherent mode which is linearly mixed into the rates of both cells. These coupled rate functions are then used to independently draw a realization of an inhomogeneous Poisson process for each cell.  Using this method 15 trials of duration 0.5s were generated.  The coherent mode was set such that the population coherence was a Gaussian of height 0.35 and standard deviation 5 Hz centered on 20 Hz.  The phase of this mode was set to $180^o$.  Figure \ref{coherency1} indicates that this coherent mode is reasonably estimated. 

\section{Example Periodic Stimulation}
\label{periodiceg}

\begin{figure}[hbtp] 
\begin{center}
\leavevmode
\hbox{%
\unitlength=1mm
\begin{picture}(0,110)
\put(-67,5){\includegraphics[height=5cm]{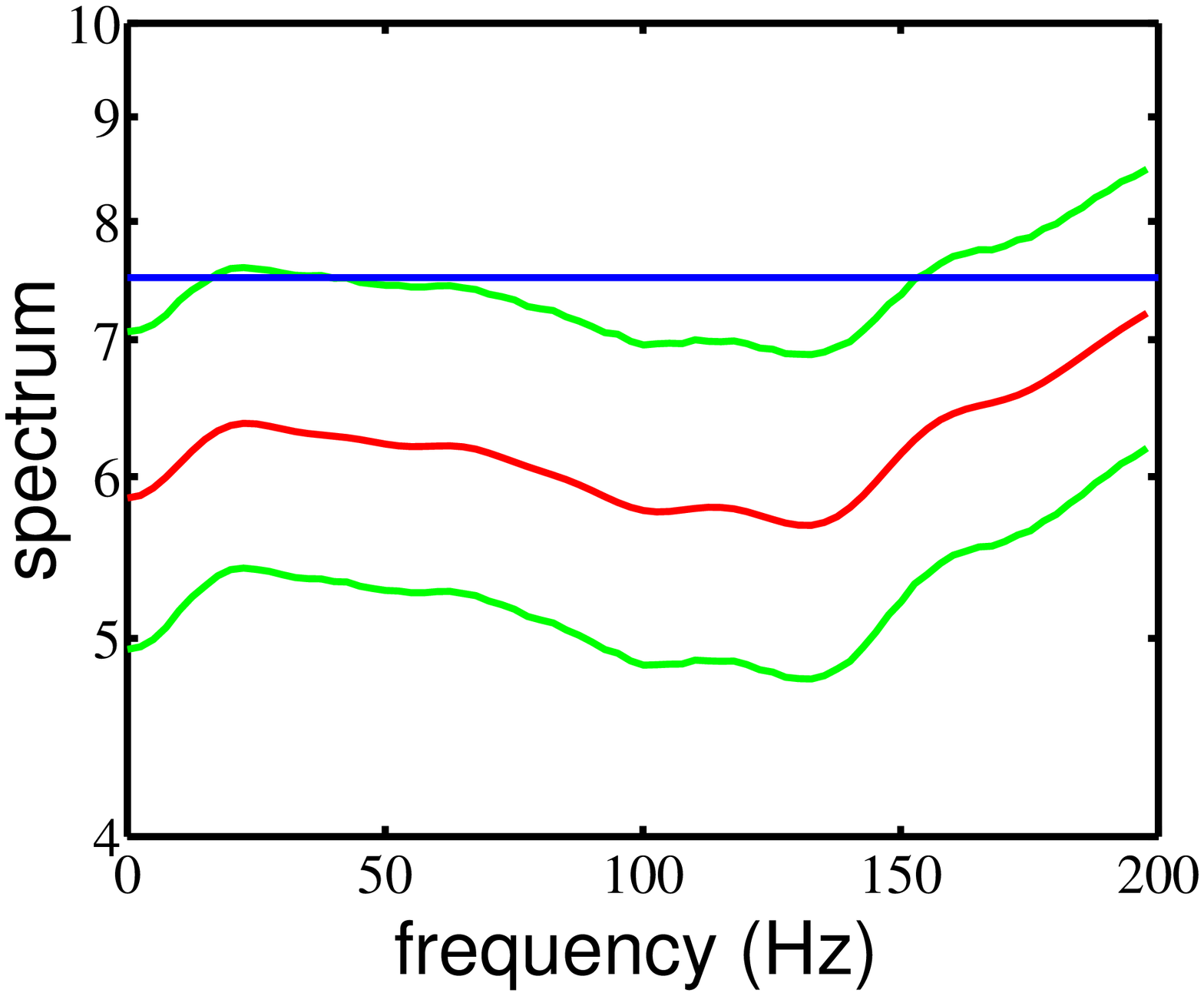}}
\put(6,5){\includegraphics[height=5cm]{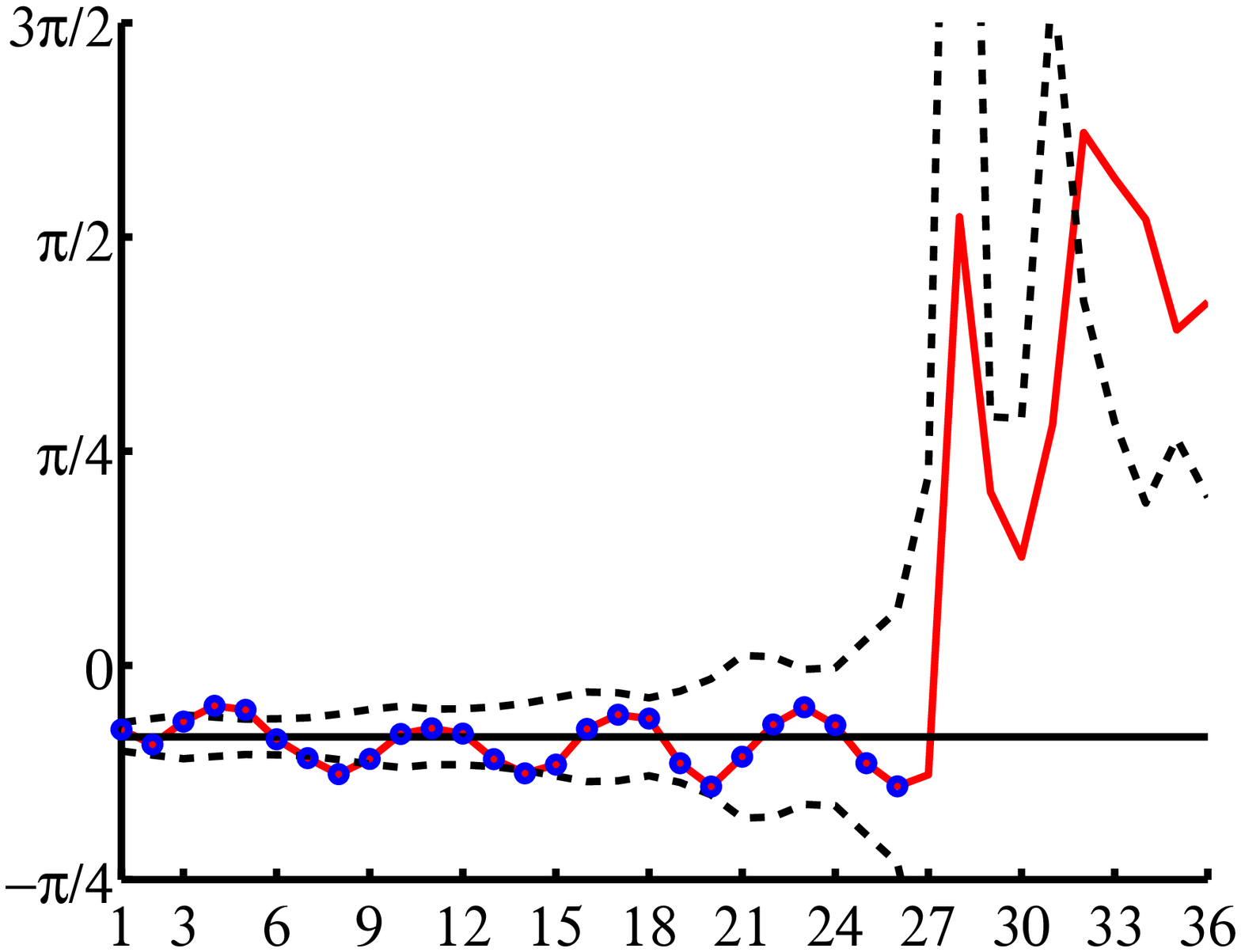}}
\put(-70,61){\includegraphics[height=5cm]{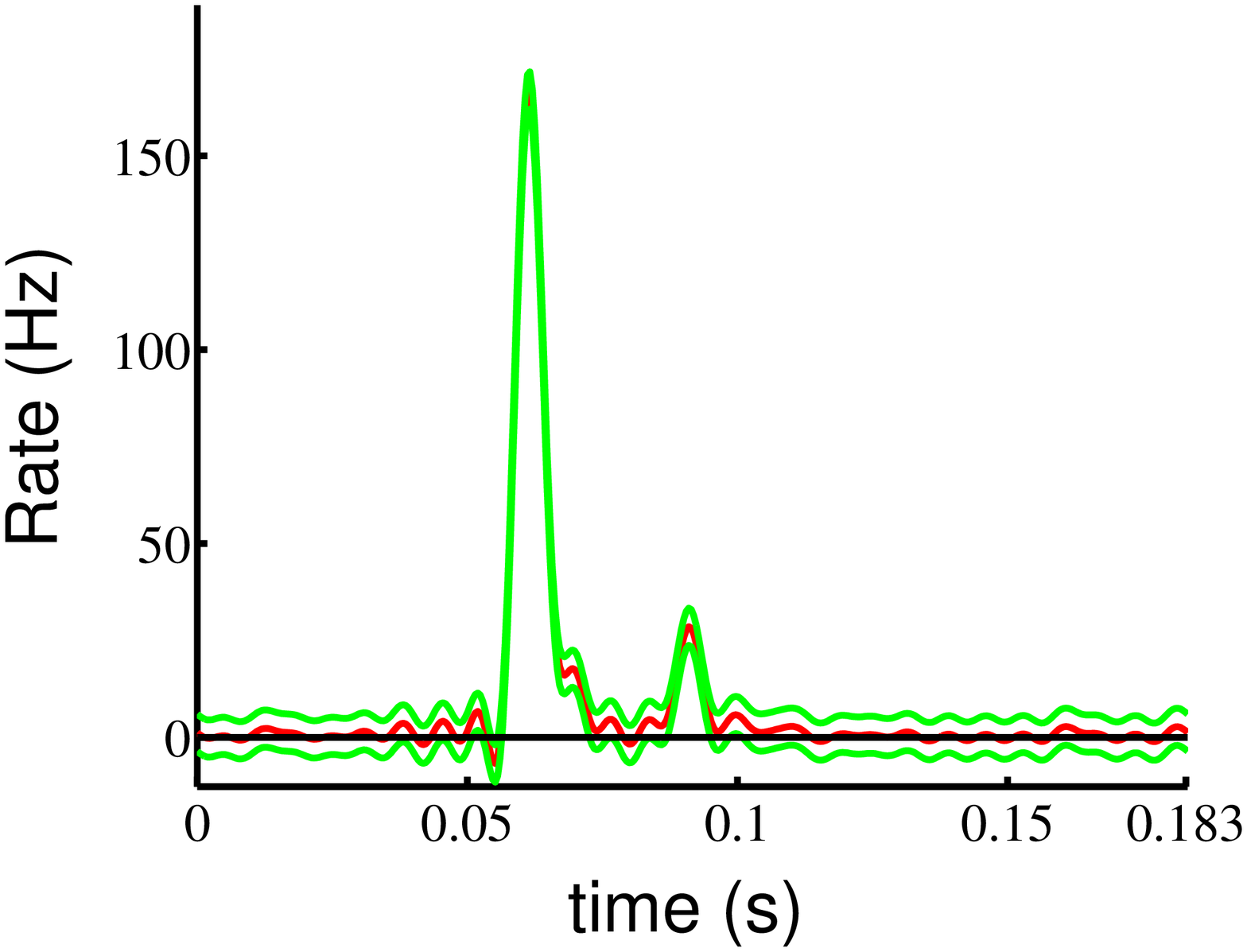}}
\put(10,64){\includegraphics[height=5cm]{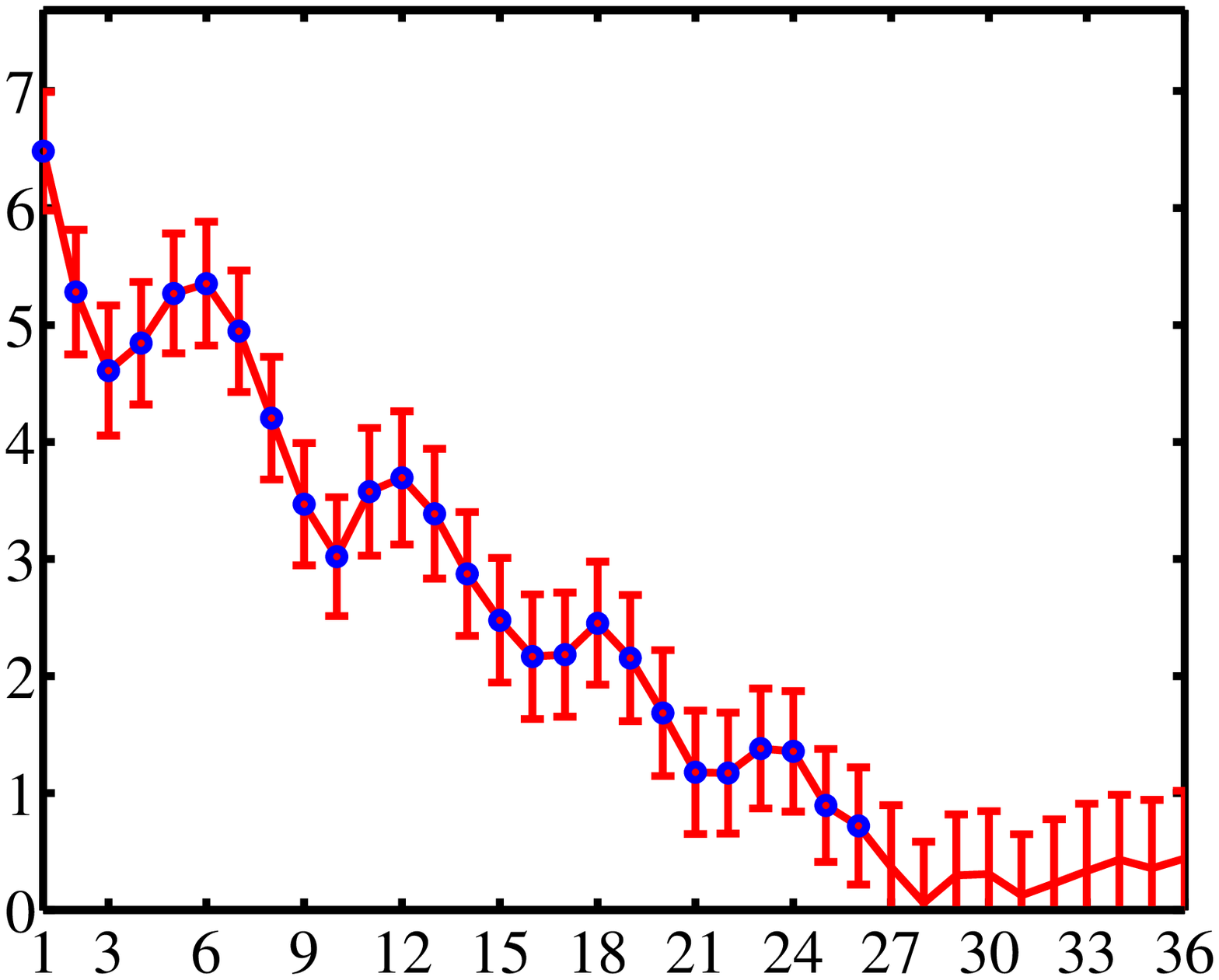}}
\put(-37,57){(a)}
\put(37,57){(b)}
\put(-37,0){(c)}
\put(37,0){(d)}
\end{picture}}
\end{center}
\caption{Response to a periodic stimulation of frequency 5.5 Hz.   (a) Impulse response function with global 95\% confidence interval (b) $| \hat c_n |$ versus index n with 95\% confidence interval.  Dots indicate points which achieved significance in the F-test. (c) Residual spectrum with finite size corrected confidence interval.  A multitaper spectrum with 100 tapers and a bandwidth of 1.5 Hz was used initially to avoid overlap of harmonics.  This spectrum was then further smoothed using a Gaussian lag window with standard deviation 9 Hz. (d) The coefficient phases $\hat \phi_n$ (in radians) versus index n after subtraction of a fitted straight line of gradient $2\pi/3 \pm 0.01$.  The black dashed lines are a 95\% confidence interval about zero.}
\label{periodicstim1}
\end{figure}

An example of an analysis of a periodic stimulus paradigm is shown in figure \ref{periodicstim1}.  The data is a single cell recording collected from the barrel cortex of an awake behaving rat during periodic whisker stimulation at 5.5 Hz \cite{Robert1}.  There is a single trial of duration 50s.    

The estimated impulse response function $\hat \lambda (t)$ is seen to have two distinct sharp peaks outside of which the response does not differ significantly from zero.  The moduli of the Fourier coefficients are significant out to $n=25$.  This automatically sets the smoothing of $\hat \lambda (t)$ as structure on a time scale of less than $1/(25\times5.5) = 7$ ms does not achieve significance.  Note that the coefficients are enhanced at multiples of 6 (i.e.$\sim33$ Hz) which comes from having two peaks in the time domain $\lambda (t)$ which are separated by $\sim30$ ms.  The phase of the coefficients closely follows a straight line but there is a small periodic deviation from this line which is again at index multiples of 6.  The gradient of the straight line depends on the time delay of the response.  The residual spectrum was calculated by first evaluating a multitaper estimate from which the significant harmonics were removed.  This spectrum had a bandwidth of 1.5 Hz chosen to avoid overlap of the harmonics leading to the multitaper estimate being undersmoothed.  A further smoothing was performed using a lag window\footnote{The previous theory developed for lag window estimators applies to this hybrid estimator with $|H(\cdot)|^2$ replaced by $\frac{1}{K} \sum_{k=0}^{K-1} |H_k(\cdot)|^2$ in equation \ref{mtbias} and $|{\cal H}(\cdot)|^2$ replaced by $\frac{1}{K} \sum_{k=0}^{K-1} |{\cal H}_k(\cdot)|^2$ in equation \ref{mtvar}.} and the resultant spectrum, displays a slight but significant suppression relative to a Poisson process out to almost 200 Hz.  Such a spectrum is characteristic of a short time scale refractive period.  The residual spectrum is particularly useful because rate non-stationarity has been removed.

\section{Summary}

It is the belief of the authors that spectral analysis is a fruitful and under exploited analysis technique for spike trains.  In this paper an attempt has been made to collect the machinery necessary for performing spectral analysis on spike train data into a single document.  Starting from the population definitions the statistical properties of estimators of the spectrum and coherency have been reviewed.  Estimation methods for both continuous spectra and spectra which contain lines have been included.  In addition new corrections to asymptotic error bars have been presented which increase confidence in applying spectral techniques in practical situations where data is often sparse.  Tables 1 to 5 summarize the important formulae. Matlab software implementing the methods discussed in this paper is available from xxx.lanl.gov/archive/neuro-sys.

\begin{table}
\begin{center}
\begin{tabular}{|c|} \hline
 $J^{kn}_{a}(f) = \int_{0}^{T} h_k(t)e^{-2\pi i ft} d\overline{N}^{n}_{a}(t)$  \\
 $I^{kn}_{ab}(f) = J^{kn}_{a}(f)J^{kn*}_{b}(f)$ \\ \hline
\end{tabular} 
\end{center}
\caption{The basic direct spectral estimator in terms of which the other estimators can be written.  For clarity the superscript $D$ on the direct spectral estimate has been omitted. The index $n$ labels trials, index $k$ labels tapers, and indices $a$ and $b$ label cells. }
\label{basicjandi}
\end{table}

\begin{table}
\begin{center}
\begin{tabular}{|c|cc|c|} \hline
$X$  & $I_{ab}^{X}(f)$  & Eq. & $\nu_0$ \\
\hline
$D$  & $I_{ab}^{01}(f) $ & [\ref{D1} ] & $2$ \\
$DT$ & $\frac{1}{N_T}\sum_{n=1}^{N_T} I^{0n}_{ab}(f) $ & [\ref{TA1} ] & $2N_T$ \\
$LW$ & $\frac{1}{N_T}\sum_{n=1}^{N_T} \int_{-\infty}^{\infty} K(f-f')I^{0n}_{ab}(f') df'$ & [\ref{LW1} ] & $2N_T/\xi$ \\
$MT$ & $\frac{1}{N_TK}\sum_{n=1}^{N_T} \sum_{k=0}^{K-1} I^{kn}_{ab}(f) $ & [\ref{MT1} ] & $2N_TK$ \\ \hline
\end{tabular} 
\end{center}
\caption{The different estimators and the large sample degrees of freedom $\nu_0$ of estimates of the spectrum ($ab=11$).  The indices on the $I_{ab}^{kn}$ are as follows.  $ab$ label the cells from which the estimates are constructed. The index $k$ labels the taper and $n$ labels the trial.} 
\label{sumallres1}
\end{table}

\begin{table}
\begin{center}
\begin{tabular}{|c|cc|c|} \hline
         & Equation  & Eq. & Comment \\ \hline 
 & & & use $\nu_0$ for asymptotic\\
Variance & $var\{I^{X}_{aa}(f)\} = \frac{2 E\{I^{X}_{aa}(f) \}^2}{\nu(f)}$ & [\ref{dof1}] & or $\nu(f)$ if using finite \vspace{-0mm} \\ & & & size correction \\ \hline  & & & \\

Degrees & $\frac{1}{\nu(f)} = \frac{1}{\nu_0} + \frac{C^X_h\Phi(f)}{2TN_TE\{I^X(f) \}^2}$ & [\ref{mainres2.5}] & See text for definitions \vspace{0mm} \\ of freedom & & & of $C^X_h$ and $\Phi(f)$ \\ \hline \ & & & \vspace{-0mm} \\

Confidence & $\left[\nu I^{X}(f)/q_2, \nu I^{X}(f)/q_1\right]$ & [\ref{conf2}] & $q_1$ s.t $P[\chi_{\nu}^{2} \leq q_1] = p$  \\$(1-2p)\times 100\%$  & & & $q_2$ s.t $P[\chi_{\nu}^{2} \geq q_2] = p$  \\ \hline

\end{tabular}
\end{center}
\caption{Main formulae required for estimating spectral error bars.  Refer to section \ref{estthespec} for additional information.}
\label{spectest1}
\end{table}

\begin{table}
\begin{center}
\begin{tabular}{|c|cc|c|} \hline  
         & Equation  & Eq. & Comment \\ \hline  & & & \\

Coherency & $C^{X}(f) = \frac{I^{X}_{ab}}{\sqrt{I^{X}_{aa}I^{X}_{bb}}}$ & [\ref{Coh0}]& \\ & & & \vspace{-3mm} \\ \hline  & & & \vspace{-0mm} \\

Distribution & $P(|C|) = (\nu-2)|C|(1-|C|^2)^{(\nu/2-2)}$ & [\ref{Nsph}] & Under null \vspace{-2mm} \\ for coherence & & & hypothesis $\gamma = 0$ \\ \hline

& & & \vspace{-0mm} \\ Confidence & $\hat{\phi}(f) \pm 2 \sqrt{\frac{2}{\nu} \left( \frac{1}{|C(f)|^2} - 1 \right)}$ & [\ref{cphsc}] & Approx. \vspace{-0mm} \\ for phase & & & $95\%$ \\ \hline

\end{tabular}
\end{center}
\caption{Main formulae required for coherency estimation.  Refer to section \ref{estthecoh} for additional information.}
\label{cohest1}
\end{table}

\begin{table}
\begin{center}
\begin{tabular}{|c|cc|c|} \hline
         & Equation  & Eq. & Comment \\ \hline  & & &  \vspace{-0mm} \\ 
Complex amplitude & $\hat c_1 = \frac{\sum_k J_k(f_1)H_k(0)}{\sum_k | H_k(0)|^2}$ & [\ref{estc1}] & \vspace{-2mm} \\ of line & & &  \\ \hline  & & & \vspace{-0mm} \\

F-test to access the & $\frac{|\hat c_1 |^2 \sum_k |H_k(0)|^2(K-1)}{\sum_k |J_k(f_1) - \hat c_1 H_k(0)|^2} \doteq F_{2,2(K-1)}$ & [\ref{fdist}]  \vspace{-2mm} & Null \\ significance of a line & & & $c_1 = 0$  \\ \hline & & & \vspace{-0mm} \\
Residual spectrum & $\hat S(f) = \frac{1}{K} \sum_k | J_k(f) - \hat c_1 H_k(f-f_1)|^2$ & [\ref{resspec}]  \vspace{-2mm} & \\ & & &  \\ \hline

\end{tabular}
\end{center}
\caption{Main formulae required for the detection and removal of a line from the spectrum.  Refer to section \ref{linspec1} for additional information.}
\label{Ftest1}
\end{table}

\pagebreak

\appendix
\section{Derivation of Finite Size Correction}
\label{app1}

The following is an outline derivation of the finite size corrections described in section \ref{fse}.   Firstly the characteristic functionals \cite{bt1} for the processes $\overline{N}$ and the inhomogeneous Poisson process rate $\lambda(t)$ are related.

\begin{equation}
C_{\overline{N}}(\theta(t)) = E\{exp (i\int_0^T \theta(t) d \overline{N}) \} = E_{\lambda} \{exp ( \int_0^T \lambda(t) b(\theta(t))dt ) \}
\end{equation}

\begin{equation}
b(\theta(t)) = exp[i\theta(t) - \frac{i}{T} \int_0^T \theta(t')dt']
\end{equation}
Under the Gaussian process assumption for $\lambda(t)$ this integral may be done.

\begin{equation}
C_{\overline{N}} = exp[\frac{1}{2} \int_0^T \int_0^T b(t)\Lambda(t,t')b(t') dtdt' + \overline{\lambda} \int_0^T b(t)dt]
\end{equation}

\begin{equation}
\Lambda(t,t') = E_{\lambda} \{ (\lambda(t) - \overline{\lambda}) (\lambda(t') - \overline{\lambda}) \}
\end{equation}
Note that $\overline{\lambda}$ denotes the mean rate.  Taking the log of the characteristic functionals now yields the following relation between the resultant cumulant functionals. 

\begin{equation}
K_{\overline{N}} = ln E\{exp (i\int_0^T \theta(t) d \overline{N}) \} = \frac{1}{2} \int_0^T \int_0^T b(t)\Lambda(t,t')b(t') dtdt' + \overline{\lambda} \int_0^T b(t)dt \label{a1}
\end{equation}
Next $\theta(t)$ is chosen appropriately and substituted into $K_{\overline{N}}$.  The form for $\theta(t)$ which is required to obtain the covariance of multitaper estimators is;

\begin{equation}
i\theta(t) = \theta_1 h_k(t)e^{-2 \pi if_1 t} + \theta_2 h_k(t)e^{2 \pi if_1 t} + \theta_3 h_{k'}(t)e^{-2 \pi if_2 t} + \theta_4 h_{k'}(t)e^{2 \pi if_2 t}
\end{equation}
Substituting into the cumulant functional for $\overline{N}$ yields;

\begin{equation}
K_{\overline{N}} = ln E\{exp(\theta_1 J^D_k(f_1) + \theta_2 J^{D*}_k(f_1) + \theta_3 J^D_{k'}(f_2) + \theta_4 J^{D*}_{k'}(f_2) ) \}
\end{equation}
Where $J^D_k$ is the Fourier transform of the data tapered by a function indexed by $k$.  Application of the cumulant expansion theorem \cite{ma1} then leads to;

\begin{equation}
K_{\overline{N}} = E\{exp(\theta_1 J^D_k(f_1) + \theta_2 J^{D*}_k(f_1) + \theta_3 J^D_{k'}(f_2) + \theta_4 J^{D*}_{k'}(f_2) ) - 1\}_C
\end{equation}
This may then be differentiated and set to zero.

\begin{equation}
K_{lmno} = \left . \frac{\partial K_{\overline{N}}}{\partial\theta_1^l \partial\theta_2^m \partial\theta_3^n \partial\theta_4^o} \right \vert_{\theta_1 = \theta_2 =\theta_3 = \theta_4 = 0} = E\{ J^{Dl}_k(f_1)J^{Dm*}_k(f_1)J^{Dn}_{k'}(f_2)J^{Do*}_{k'}(f_2) \}_C
\end{equation}
Moments of the estimators may be expressed in terms of these cumulant derivatives.  The expressions are simplified by the fact that all cumulant derivatives which have indices summing to an odd number are zero because $\overline{N}$ is a zero mean process.  
\begin{equation}
E\{ I^D_k(f)\} = K_{1100}
\end{equation}

\begin{equation}
var\{I^{MT}(f)\} = \frac{1}{K^2} \sum_{k=0}^{K-1} \sum_{k'=0}^{K-1} cov\{ I^D_k(f),I^D_{k'}(f) \} 
\end{equation}

\begin{equation}
cov\{ I^D_k(f),I^D_{k'}(f) \} = K_{1010}K_{0101} + K_{1111} + K_{1001}K_{0110} 
\end{equation}
The problem has now been reduced to that of calculating these derivatives within the model.  This is done by substituting the expression for $\theta(t)$ into the RHS of equation \ref{a1}.  Considerable algebra then leads to the following exact result.

\begin{equation} 
K_{lmno} = K_{lmno}^A + K_{lmno}^B
\end{equation}
Where,
\begin{eqnarray}
K_{lmno}^A = \frac{1}{2} \sum_{l_i,m_i,n_i,o_i} \frac{l!m!n!o!}{\Pi l_i! \Pi m_i!  \Pi n_i!  \Pi o_i!} 
\left [ \frac{-H_1(f_1)}{T} \right ]^{l_2 + l_4}
\left [ \frac{-H_1(f_1)^*}{T} \right ]^{m_2 + m_4}\hspace{0cm} ... \nonumber \\
\left [ \frac{-H_1(f_2)}{T} \right ]^{n_2 + n_4}
\left [ \frac{-H_1(f_2)^*}{T} \right ]^{o_2 + o_4}
I^{l_1,m_1,n_1,o_1}_{l_3,m_3,n_3,o_3} 
\end{eqnarray}
Where $ \sum_i l_i = l$ and cases where $l_1 + l_2 = l$ or $l_3 + l_4 = l$ are excluded (and also for $n,m,o$).
\begin{eqnarray}
I^{l_1,m_1,n_1,o_1}_{l_3,m_3,n_3,o_3} = \int_{\infty}^{\infty} S_{\lambda} (f) H_{l_1+m_1+n_1+o_1}[f_1(l_1-m_1)+f_2(n_1-o_1) - f] \hspace{1cm} ... \nonumber \\ 
H^*_{l_3+m_3+n_3+o_3}[f_1(l_3-m_3)+f_2(n_3-o_3) - f]df
\end{eqnarray}
Where $S_{\lambda}(f)$ is the spectrum of the Gaussian process and $H_l$ is;
\begin{eqnarray}
H_l(f) = \int_{-\infty}^{\infty} h^l(t)exp(-2 \pi i f t) dt \\
H_0(f) = Texp(-i \pi fT) sinc(\pi fT)
\end{eqnarray}
\begin{eqnarray}
K^B_{lmno} = \overline{\lambda} \sum_{p=0}^{l}\sum_{q=0}^{m}\sum_{r=0}^{n}\sum_{s=0}^{o}
[ \begin{array}{l} l \vspace{-4mm} \\  p \end{array}  ] 
[ \begin{array}{l} m \vspace{-4mm} \\ q \end{array}  ]
[ \begin{array}{l} n \vspace{-4mm} \\ r \end{array}  ]
[ \begin{array}{l} o \vspace{-4mm} \\ s \end{array}  ]
H_{p+q+r+s} [f_1(p-q) + f_2(r-s)] \hspace{2mm}  ... \nonumber \\
\left [ \frac{-H_1(f_1)}{T}   \right ]^{(l-p)}
\left [ \frac{-H_1(f_1)^*}{T} \right ]^{(m-q)}
\left [ \frac{-H_1(f_2)}{T}   \right ]^{(n-r)}
\left [ \frac{-H_1(f_2)^*}{T} \right ]^{(o-s)}
\end{eqnarray}
The preceding result is somewhat cumbersome but readily evaluated computationally for a given spectrum.  The expression simplifies greatly when only frequencies above the smoothing width are considered and many of the terms may be neglected.  Restricting attention to the second order properties there are only a few remaining dominant terms.  Terms from $K_{1001}$ lead to the previously discussed asymptotic results but there are corrections which arise from the term $K_{1111}$.  Assuming that the population spectrum varies slowly over the width of the tapers leads to the result given by equations \ref{mainres2} - \ref{mainres6}.  The validity of this assumption has been tested computationally and was found to be very accurate even for spectra with sharp peaks.

\normalsize

\section*{Acknowledgment} The authors thank C. Buneo, and R. Sachdev for providing example datasets, Clive Lauder for help with the calculation of global error bars and D. R. Brillinger and D. J. Thomson for comments which substantially improved the manuscript.  M. Jarvis is grateful to R. A. Andersen both for his continued support of theoretical work in his lab and also for his careful reading of the manuscript. M. Jarvis acknowledges the generous support of the Sloan foundation for theoretical neuroscience.

\bibliography{jarvis_et_al}
\end{document}